\documentclass[fleqn,10pt]{wlscirep}
\usepackage[utf8]{inputenc}
\usepackage[T1]{fontenc}
\usepackage{comment}
\usepackage{amsmath}
\usepackage{float}

\usepackage{upgreek}
\graphicspath{{images/}}
\usepackage{tikz}

\newcommand{\thetaB}{\uptheta}
\newcommand{\eB}{\Vec{e}}
\newcommand{\rB}{\vec{r}}
\newcommand{\RB}{R}
\newcommand{\kB}{\vec{k}}
\newcommand{\mB}{\textbf{M}}
\newcommand{\nBB}{\textbf{n}}
\newcommand{\nB}{\vec{n}}
\newcommand{\bB}{\textbf{B}}
\newcommand{\xiB}{\vec{\xi}}
\DeclareMathSymbol{\shortminus}{\mathbin}{AMSa}{"39}
\usepackage{graphicx}
\usepackage{algorithm,algorithmic}
\graphicspath{{images/}}

\title{Real space iterative reconstruction for vector tomography (RESIRE-V)}

\author[1,2,3,*]{Minh Pham}
\author[1,4]{Xingyuan Lu}
\author[1]{Arjun Rana}
\author[2,3]{Stanley Osher}
\author[1,*]{Jianwei Miao}

\affil[1]{Department of Physics \& Astronomy and California NanoSystems Institute, University of California, Los
Angeles, CA 90095, USA}
\affil[2]{Department of Mathematics, University of California, Los Angeles, CA 90095, USA}
\affil[3]{Insitute of Pure and Applied Mathematics, University of California, Los Angelese, CA 90095, USA}
\affil[4]{School of Physical Science and Technology, Soochow University, Suzhou 215006, China}
\affil[*]{minhrose@ucla.edu; miao@physics.ucla.edu}

\begin{abstract}
Tomography has had an important impact on the physical, biological, and medical sciences. To date, most tomographic applications have been focused on 3D scalar reconstructions. However, in some crucial applications, vector tomography is required to reconstruct 3D vector fields such as the electric and magnetic fields. Over the years, several vector tomography methods have been developed. Here, we present the mathematical foundation and algorithmic implementation of REal Space Iterative REconstruction for Vector tomography, termed RESIRE-V. RESIRE-V uses multiple tilt series of projections and iterates between the projections and a 3D reconstruction. Each iteration consists of a forward step using the Radon transform and a backward step using its transpose, then updates the object via gradient descent. Incorporating with a 3D support constraint, the algorithm iteratively minimizes an error metric, defined as the difference between the measured and calculated projections. The algorithm can also be used to refine the tilt angles and further improve the 3D reconstruction. To validate RESIRE-V, we first apply it to a simulated data set of the 3D magnetization vector field, consisting of two orthogonal tilt series, each with a missing wedge. Our quantitative analysis shows that the three components of the reconstructed magnetization vector field agree well with the ground-truth counterparts. We then use RESIRE-V to reconstruct the 3D magnetization vector field of a ferromagnetic meta-lattice consisting of three tilt series. Our 3D vector reconstruction reveals the existence of topological magnetic defects with positive and negative charges. We expect that RESIRE-V can be incorporated into different imaging modalities as a general vector tomography method.
To make the algorithm accessible to a broad user community,, we have made our RESIRE-V MATLAB source codes and the data freely available at \url{https://github.com/minhpham0309/RESIRE-V}.
\end{abstract}

\begin{document}

\flushbottom
\maketitle
%
%
\thispagestyle{empty}

\section*{Introduction}

Tomography has had a radical impact on diverse fields ranging from medical diagnosis\cite{kak2002} to 3D structure determination of proteins\cite{frank2006}, crystal defects\cite{Scott2012, Miao2016} and amorphous materials\cite{Yang2021, Yuan2022}, at the atomic resolution. Despite its very diverse applications, the central problem in tomography remains the same, that is, how to accurately reconstruct the 3D structure of an object from a number of projections with noise and incomplete data. The conventional reconstruction methods include filtered back projection (FBP)\cite{kak2002, frank2006}, algebraic reconstruction technique (ART)\cite{gordon1970}, simultaneous algebraic reconstruction technique (SART)\cite{andersen1984}, and simultaneous iterative reconstruction technique (SIRT)\cite{gilbert1972, herman2009}, which remain popular in tomographic applications. Recently, more advanced iterative algorithms have been developed for tomography, including equal slope tomography (EST)\cite{miao2005, lee2008}, nonuniform fast Fourier transform (NUFFT)\cite{oconnor2006}, generalized Fourier iterative reconstruction (GENFIRE)\cite{yang2017, pryor2017} and real space iterative reconstruction (RESIRE)\cite{Yang2021, Pham_resire}. In particular, RESIRE, which uses the Radon transform as the forward projection and the Radon transpose as the back projection, is not only superior to other existing tomographic algorithms\cite{Pham_resire}, but also has been used to determine the 3D atomic structure of amorphous materials\cite{Yang2021, Yuan2022} and the chemical order and disorder in medium/high entropy alloys\cite{moniri2023}. Despite all these applications, they only deal with scalar tomography, where each voxel in a 3D reconstruction has a magnitude but no direction. However, in some important applications, vector tomography is required, where each voxel has a magnitude and a direction such as the electric and magnetic field. Over the years, several vector tomography reconstruction methods have been developed, including vector electron tomography with Lorentz transmission electron microscopy and holography\cite{PHATAK2008, Phatak2010, Phatak2016, wolf2019, yu2020, wolf2022, lewis2022a, lewis2022b}, soft and hard x-ray vector tomography\cite{Streubel2015, Donnelly2017, donnelly2020, hierro2020, witte2020, donnelly2021, donnelly2022, hermosa2022,  rana_2023, raftrey2023 }. In particular, the combination of ptychography, a powerful coherent diffractive imaging method\cite{miao1999, miao2015}, and vector tomography can in principle achieve the highest spatial resolution, which is only limited by the wavelength and the diffraction signal\cite{Donnelly2017, donnelly2021, rana_2023}. Very recently, we have merged soft-x-ray magnetic circular dichroism and ptychography with vector tomography to image the 3D topological magnetic monopoles and their interaction in a ferromagnetic meta-lattice with a spatial resolution of 10 nm\cite{rana_2023}. Using RESIRE\cite{Pham_resire}, our vector tomography algorithm can accurately reconstruct the 3D magnetization vector field from multiple tilt series each with a limited number of experimental projections. Furthermore, due to the experimental error, the measured tilt angles may not always coincide with the true orientations of the projections. To tackle this problem, we implement an iterative angular refinement method to reduce the tilt angle error\cite{Pham_resire}, enabling us to obtain more accurate vector tomographic reconstruction. Here, we provide the mathematical foundation and implementation of our vector tomography algorithm, termed RESIRE-V. Both numerical simulations and experimental data have been used to demonstrate the effectiveness of this vector tomography algorithm.

\section*{Methods}
We begin with some setup and conventions. First, we employ Euler angles to describe the orientation of a rigid body with respect to a fixed coordinate system. 
For example, the orientation representation ZYX used intensively in our research fits well with vector tomography experiments: samples are rotated about the Z-axis (in-plane rotation) before a set of tilt series (rotation about the Y-axis) are acquired. The last rotation about the X-axis is helpful in angular refinement. 
We use the notation $Z_{\phi}Y_{\theta}X_{\psi}$ to represent Euler angle rotations: the first rotation is about the Z-axis by an angle $\phi$, followed by a rotation about the Y-axis by an angle $\theta$, and ends with a rotation about the X-axis by an angle $\psi$, respectively.
The corresponding rotation matrix $\RB_{Z_{\phi}Y_{\theta}X_{\psi}} =  \RB_{\phi}^Z \, \RB_{\theta}^Y \, \RB_{\psi}^X$ is defined to be the product of three single-axis rotation matrices about the Z, Y, and X axes by angles $\psi$, $\theta$ and $\phi$ respectively:
\begin{align*}
\RB_{\phi}^Z := 
    \begin{bmatrix}
    \cos \phi & -\sin \phi & 0\\
    \sin \phi & \cos \phi & 0 \\
    0 & 0 & 1
    \end{bmatrix}
    ,\quad \RB_{\theta}^Y := 
    \begin{bmatrix}
    \cos \theta & 0 & \sin \theta \\
    0 & 1 & 0 \\
    -\sin \theta  & 0 &\cos \theta
    \end{bmatrix}
    , \quad  \RB_{\psi}^X := 
    \begin{bmatrix}
    1 & 0 & 0 \\
    0 &  \cos \psi & -\sin \psi \\
    0 &  \sin \psi &  \cos \psi
    \end{bmatrix}
\end{align*}
For short notation, we write $\RB_{\thetaB}$ instead of $\RB_{Z_{\phi}Y_{\theta}X_{\psi}}$ where $\thetaB = \{ \phi, \theta, \psi \}$ (no orientation is specified). 
In perfect experimental conditions where there is no X-axis rotation, $\psi$ is zero. 
Otherwise, $\psi$ can be non-zero and we use angular refinement to determine $\psi$. The convention finishes and we move to the formulation part.

\subsection*{Formulation}

For an x-ray beam propagating along the z direction (standard unit vector $\eB_z = [0, \; 0, \; 1]^T$), only the z component of the magnetization contributes to the 2D signals. The contribution takes either positive or negative values depending on the left or right circular polarization. In the case of rotation, we need the inner product $\big\langle \RB_{\thetaB} \; \mB (\RB_{\thetaB}^{\dagger} \rB )
    , \;  \kB \big\rangle$ to count for the contribution.
Here, $\mB = [M_x, \; M_y, \; M_z ]$ is the magnetization vector field, which is a function of the Cartesian coordinate vector $\rB = ( x, \; y, \; z )$, and $ \RB_{\thetaB}^\dagger$ is the adjoined, and also inverse and transpose, of $ \RB_{\thetaB}$.
Adding the non-magnetic term $O$ and taking the integral along the z axis (projection), we obtain the 2D signal:
\begin{align}
    \int_z  c \big\langle \RB_{\thetaB} \; \mB (\RB_{\thetaB}^\dagger \rB ) ,
    \eB_z \big\rangle + O (\RB_{\thetaB}^{\dagger} \rB)\, dz = P_{\thetaB}
\end{align}
where $c$ is a constant that relates the XMCD signal to the magnetization and the pixel size. We can temporarily let $\mB$ absorb $c$ in the derivation for simplicity and then rescale $\mB$ after the reconstruction.
We then write this equation using the change of variable $\rB \leftarrow \RB_{\thetaB}^{\dagger} \rB$:
\begin{align}
    \int_{L_{\thetaB}}  \big\langle  \mB (x,y,z) ,
    \RB_{\thetaB}^{\dagger}  \eB_z \big\rangle + O ( x,y,z )\, dz = P_{\thetaB} (x,y)
\end{align}
Rotating the sample by some Euler angles $\thetaB$ and taking the integral along the z-axis is equivalent to taking the line integral along the opposite rotation direction (passive rotation).
To solve this equation numerically, we need to discretize the equation. Replacing the line integral with a projection operator and expanding the inner product, we represent the equation algebraically:
\begin{align}\label{eqn1} 
    \Pi_{\thetaB} \Big( \alpha_{\thetaB}  M_x + \beta_{\thetaB} M_y + \gamma_{\thetaB} \,  M_z + O \Big) = P_{\thetaB}    
\end{align}
where $\Pi_{\thetaB}$ is the projection operator, and $P_{\thetaB}$ is the corresponding projection with respect to Euler angles $\thetaB = (\phi, \;  \theta, \; \psi)$. In this notation, we drop the spatial variables $(x,y,z)$ for simplicity. Let $ \nB_{\thetaB} = [\alpha_{\thetaB}, \; \beta_{\thetaB}, \; \gamma_{\thetaB}]$ be the last column of $\RB_{\thetaB}^\dagger$. Specifically, if we use the orientation representation $Z_{\phi}Y_{\theta}$, then the normal vector is given by:    $\nB_{\thetaB} = [\alpha_{\thetaB}, \; \beta_{\thetaB}, \; \gamma_{\thetaB}] = [\sin\theta \, \cos\phi, \; \sin\theta  \, \sin\phi  , \; \cos\theta]$.

One can verify that the second magnetization component does not contribute to the measured projections when $\phi=0$. It implies that other types of rotation are required for successful vector tomography reconstructions. Since $\Pi_{\thetaB}$ is linear, we can apply the commutative property and distribute the linear operator to each magnetization component:
\begin{align}\label{eqn_full}
    \alpha_{\thetaB} \, \Pi_{\thetaB}  (M_x) + \beta_{\thetaB} \, \Pi_{\thetaB} (M_y) + \gamma_{\thetaB} \, \Pi_{\thetaB}  (M_z) +\Pi_{\thetaB} (O) = P_{\thetaB}
\end{align}
Eqn.\ref{eqn_full} describes that the three 3D magnetization components and the non-magnetic structure are coupled via a linear constraint.
So far, we have formulated the vector tomography in the perfect condition (noise free). 
In the presence of noise and assuming that the noise is Gaussian with mean $0$ and variance $\sigma^2$, we add the noise term $\mathcal{N}(0,\sigma^2)$ to the left-hand side of the equation.
\begin{align}\label{eqn_noise}
    \alpha_{\thetaB} \, \Pi_{\thetaB} (M_x) + \beta_{\thetaB} \, \Pi_{\thetaB} (M_y) + \gamma_{\thetaB} \, \Pi_{\thetaB} (M_z) +\Pi_{\thetaB} (O) + \mathcal{N}(0,\sigma^2) = P_{\thetaB} 
\end{align}
We first denote $P_{\thetaB}^+$ and $P_{\thetaB}^-$ are random variables, with the same variance $\sigma^2$, that represent the left and right polarized projections by Euler angles $\thetaB$ respectively. Let $b_{\thetaB}^- =  \frac{1}{2} \big( P_{\thetaB}^+ - P_{\thetaB}^- \big)$ be a random variable as describe, we obtain a simpler linear equation:
\begin{align}\label{eqn_noise_final}
    \alpha_{\thetaB} \, \Pi_{\thetaB} (M_x) + \beta_{\thetaB} \, \Pi_{\thetaB} (M_y) + \gamma_{\thetaB} \, \Pi_{\thetaB} (M_z)  + \mathcal{N}(0,\frac{\sigma^2}{2}) = b_{\thetaB}^- 
\end{align}
Note that, by the law of large numbers, taking the average of two random variables with the same mean and variance results in a new random variable where the mean stays the same, but the variance gets reduced by half \cite{lemons2003}. We can use maximum likelihood estimation to recover three-dimensional magnetization from corrupted 2D signals. Specifically for Gaussian noise, the log maximum likelihood function is the sum of the squared errors (or $l_2$ distances) between the desired  and measured signals:
\begin{align}\label{min1}
    \min_{\mB} \; \varepsilon(\mB) &=  \frac{1}{2}\sum_{\thetaB} { \big\| \alpha_{\thetaB} \, \Pi_{\thetaB} (M_x) + \beta_{\thetaB} \, \Pi_{\thetaB} (M_y) + \gamma_{\thetaB} \, \Pi_{\thetaB} (M_z)  - b_{\thetaB}^-  \big\|^2 }
\end{align}
We can always write the minimization problem in the form $\varepsilon(\mB) = \frac{1}{2}\sum_{\thetaB} { \big\| \Pi_{\thetaB} \big( \alpha_{\thetaB} M_x + \beta_{\thetaB}  \, M_y + \gamma_{\thetaB}  \, M_z \big)  - b_{\thetaB}^-  \big\|^2 }$ thanks to the linearity of the projection operator. For efficient implementation, the latter form is preferred over Eqn. \ref{min1}.

The maximum likelihood function will appear different for other types of noise; however, the famous least-squares form can still handle other circumstances because of its simplicity and effectiveness. Eqn.\ref{min1} is our final form of the vector tomography formulation, and the remaining part is designing a numerical scheme to solve this minimization.

\subsection*{RESIRE-V algorithm}
We develop our algorithm based on the real space iterative technique.
Noticing that the projection operator is linear, one can construct a matrix representation for each $\Pi_{\thetaB}$. 
We assume that the projections have size $n\times n$, and the sample gets reconstructed with thickness $n$. In that case, each projection matrix $\Pi_{\thetaB}$ has $O(n^3)$ non-zero elements. 
In the case of over-constraint, Eqn. \ref{min1} can be solved using the normal equation. 
Otherwise, in the case of under-determined system, we need to add a regularizer to prevent overfitting. When adding a damping term as a regularizer, we have an overconstrainted system again and we can solve the equation using the normal equation as in the over-constraint case.
In either case, storing projection matrices is tremendously expensive since the size expands in the cubic order of the projection size. Here, to save memory usage, we do not need to store the projection matrices but compute the forward projections at every iteration instead. This procedure will increase the number of computations; however, GPU parallel computing can help reduce the computational time significantly.

\begin{figure}
    \centering
    \includegraphics[width=6cm]{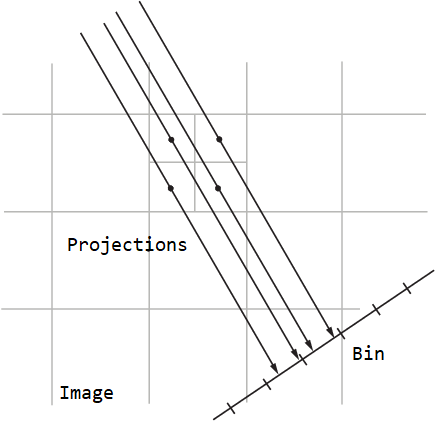}
    \caption{Illustration of the Radon transform in the 2D case from Matlab\cite{matlab_radon}: The algorithm first divides image pixels into four sub-pixels and projects them onto a 2D plane separately. The value of each sub-pixel is distributed proportionally to two nearest neighbors, according to the distance between the projected location and the pixel centers. 
    The transpose of the Radon transform follows the same idealogy in the reverse order.
    According to the distance between the projected location and the pixel centers, the two nearest neighbors to a projection sub-pixel proportionally contribute their values to the sub-pixel.}
    \label{fig:Radon}
\end{figure}

Our gradient descent algorithm incorporates two steps: forward projection and back projection. For the first step, we institute our 3D Radon transform from 2D Radon transform (Fig. \ref{fig:Radon}),
which can be found elsewhere \cite{bracewell_1995, lim_1990}.
The algorithm first divides pixels in a 3D image into four sub-pixels and projects each sub-pixel individually. Specifically, at a tilt angle, we compute the corresponding coordinate of each pixel and project it on the XY plane. The value of each sub-pixel is distributed proportionally to the four nearest neighbors, according to the distance between the projected location and the pixel centers. Supposing that the pixel projection lands on the center point of a bin, then the bin on the axis gets the entire value of the pixel. If the pixel projection hits the border between four bins, the pixel value is split evenly between these four bins. 

Next, we establish the transpose of the Radon transform for the back-projection step. This process is similar to the forward projection but in reverse order. According to the distance between the projected location and the pixel centers, the four nearest neighbors to a projection sub-pixel proportionally contribute their values to the sub-pixel. If the pixel projection hits the border between 4 bins, the pixel takes a quarter value of each of these four bins.

After specifying the forward and back projection, we can now take the gradient of the error metric  $\varepsilon(\mB)$ in Eqn. \ref{min1} with respect to each magnetization component:
\begin{align}\label{grad_mx}
    \frac{\partial \varepsilon}{\partial M_x} = \sum_{\thetaB} \alpha_{\thetaB} \, \Pi_{\thetaB}^T \, \Big( \alpha_{\thetaB} \, \Pi_{\thetaB} (M_x) + \beta_{\thetaB} \, \Pi_{\thetaB} (M_y) + \gamma_{\thetaB} \, \Pi_{\thetaB} (M_z)  - b_{\thetaB}^-  \Big) 
     = \sum_{\thetaB} \alpha_{\thetaB} \, \Pi_{\thetaB}^T \Big( \Pi_{\thetaB} \big( \alpha_{\thetaB} M_x + \beta_{\thetaB}  \, M_y + \gamma_{\thetaB}  \, M_z \big)  - b_{\thetaB}^- 
 \Big)
\end{align}
where $\Pi_{\thetaB}^T$ is the transpose operator of the Radon transform for Euler angles $\thetaB$. As mentioned above, the second form of the gradient will be used for the C++/Cuda implementation.

\begin{figure}
    \centering
    \includegraphics[width=14cm]{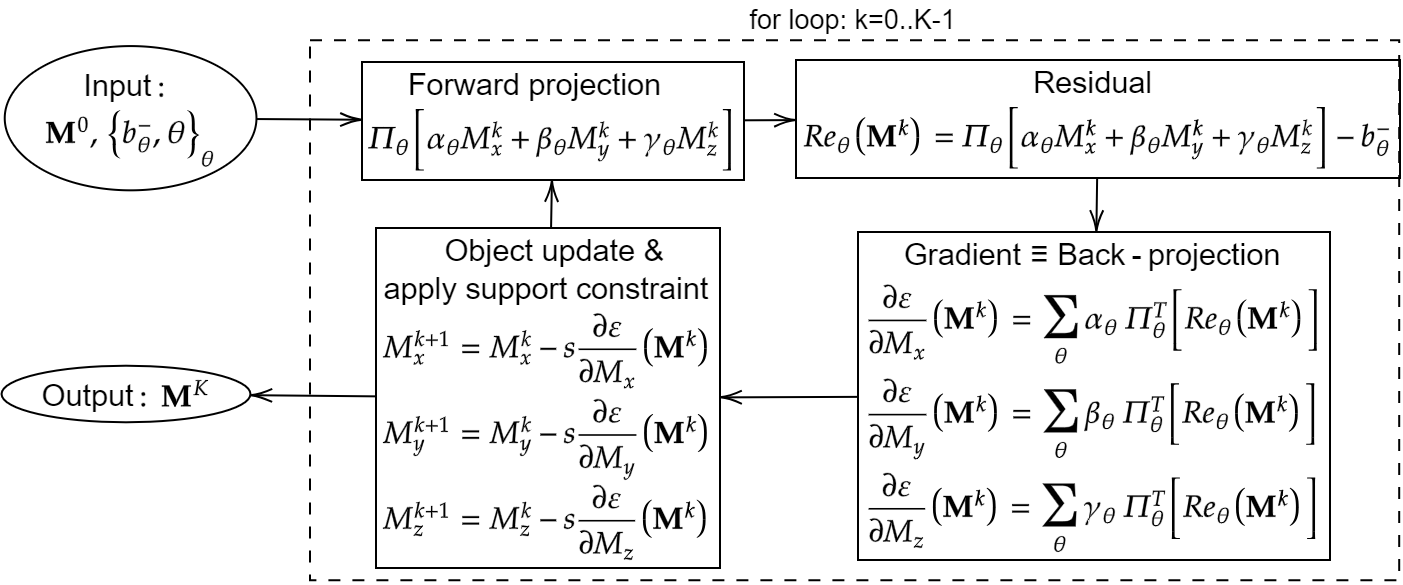}
    \caption{RESIRE-V diagram: Inputs are the differences between the left and right polarization projection and the support from the scalar reconstruction. The algorithm uses a for loop to refine the magnetization vector field $\mB$. At each iteration, it calculates the forward projections and computes their differences with the measured ones. The residuals (or differences) are back-projected to yield gradients. The algorithm will use these gradients to update the magnetization and apply the support constraint. The step size $ \frac{t} {\sqrt{3}nN}$ is replaced by $s$ for simplification purposes.}
    \label{fig:algorithm_diagram}
\end{figure}

Next, we show that the gradient is L-lipschitz and the algorithm will converge to the global minimum with an appropriate step size. Specifically, we want to find an L such that the following inequality is true:
\begin{equation}
    \big\| \nabla \varepsilon(\mB_1) - \nabla \varepsilon(\mB_2) \big\| \le L \big\| \mB_1 - \mB_2 \big\| \quad \forall  \; \mB_1, \, \mB_2
\end{equation}
The Lipchitz constant L gets calculated as $\sqrt{3} n N_z$ where $n$ and $N_Z$ are the number of projections and the thickness in pixels of the reconstruction, respectively. Hence, we can choose the step size to be $1/L$ for the convergence guarantee. Details of the proof can be found in the Supplementary, step-size analysis. 
The algorithm is finalized and described step by step in  pseudocode \ref{alg1} and Fig. \ref{fig:algorithm_diagram}.

\begin{algorithm}
\caption{RESIRE-V}
\label{alg1}
\textbf{Input}: a set of $n$ projections with left and right polarization $\{P_{\thetaB_i}^+\}_{i=1}^n$ and $\{P_{\thetaB_i}^-\}_{i=1}^n$ and their corresponding tilt angles $\{\thetaB_i\}_{i=1}^n$, the total number of iterations $K$, the step size $t \approx 1$ \\
\textbf{Preprocessing}:

    \hspace{0.5cm} 1. Compute the mean images from the left and right polarization projections: $b_{\thetaB_i}^+ = \frac{1}{2} (P_{\thetaB_i}^+ + P_{\thetaB_i}^-)$. 
    
    \hspace{0.5cm} 2. Reconstruct the non-magnetic signal $O$ from the mean images and use threshold to obtain its support.

    \hspace{0.5cm} 3. Compute the difference images from the left and right polarization projections: $b_{\thetaB_i}^- = \frac{1}{2} (P_{\thetaB_i}^+ - P_{\thetaB_i}^-)$\\
\textbf{Initialize}: $O^0$.
\begin{algorithmic}
\FOR {$k = 0,\dots,K-1$}
\FOR {$i=1,\dots,n$} 
\STATE Compute ``forward projections'' $\Pi_{\thetaB_i} (\alpha_{\thetaB_i} M_x^k + \beta_{\thetaB_i} M_y^k + \gamma_{\thetaB_i}M^k_z)$  using  Radon transform.\\

\STATE Compute residual
$$Re_{\thetaB_i}(\mB^k) := \Pi_{\thetaB_i} (\alpha_{\thetaB_i} M_x^k + \beta_{\thetaB_i} M_y^k + \gamma_{\thetaB_i}M^k_z)  - b_{\thetaB_i}^-$$

\STATE Compute ``back projection'' for each projection difference
\begin{align*}
   \frac{\partial \varepsilon_i}{\partial M_x}(\mB^k) :=  \alpha_{\thetaB_i} \, \Pi_{\thetaB_i}^T  \; Re_{\thetaB_i}(\mB^k) ,\quad
   \frac{\partial \varepsilon_i}{\partial M_y}(\mB^k) :=  \beta_{\thetaB_i} \, \Pi_{\thetaB_i}^T  \; Re_{\thetaB_i}(\mB^k) ,\quad
   \frac{\partial \varepsilon_i}{\partial M_z}(\mB^k) :=  \gamma_{\thetaB_i} \, \Pi_{\thetaB_i}^T  \; Re_{\thetaB_i}(\mB^k)
\end{align*}
\ENDFOR \\
\textbf{Update $\mB^{k+1}$}: 
\begin{align*}
    M_x^{k+1} = M_x^k - \frac{t}{\sqrt{3}nN} \frac{\partial \varepsilon}{\partial M_x} (\mB^k) \quad \text{where} \quad \frac{\partial \varepsilon}{\partial M_x}(\mB^k) := \sum_{i=1}^n \frac{\partial \varepsilon_i}{\partial M_x} (\mB^k) \\
    M_y^{k+1} = M_y^k - \frac{t}{\sqrt{3}nN} \frac{\partial \varepsilon}{\partial M_x} (\mB^k) \quad \text{where} \quad \frac{\partial \varepsilon}{\partial M_y} (\mB^k) := \sum_{i=1}^n \frac{\partial \varepsilon_i}{\partial M_y} (\mB^k) \\
    M_z^{k+1} = M_z^k - \frac{t}{\sqrt{3}nN}  \frac{\partial \varepsilon}{\partial M_x} (\mB^k) \quad \text{where} \quad \frac{\partial \varepsilon}{\partial M_z} (\mB^k) := \sum_{i=1}^n \frac{\partial \varepsilon_i}{\partial M_z} (\mB^k)
\end{align*}
Apply support constraint and other regularizers if applicable.
\ENDFOR
\end{algorithmic}
\textbf{Output}: $\mB^K$ 
\end{algorithm}

For efficient implementation, the gradient w.r.t. each component $ \frac{ \partial\varepsilon } {\partial M_x} = \sum_{i} \frac{\partial\varepsilon_i} {\partial M_x}$ will be accumulated.
In addition, the step size is generalized to be $ \frac{t} { \sqrt{3} n N} $ where $t \approx 1$ is the normalized step size. By our analysis, $t$ should be less than or equal to 1 for the convergence guarantee. The analysis uses triangle inequalities and takes into account the worst-case scenario. In practice where better scenarios are more popular, the algorithm can converge with $t$'s values slightly larger than 1.  
The analysis is complete and we move to the discussion on conditions for vector tomography reconstruction.

\subsection*{Analysis: conditions for vector tomography reconstruction}
Scalar tomography, a well-posed problem, only requires one dataset (single tilt axis) for reconstruction\cite{Rosier1968}, assuming the number of measurements is sufficient. The Fourier slice theorem can verify that a 2D image needs $n$ sampling points on the Fourier domain for a unique reconstruction. This requirement implies that $n^2$ projections corresponding to these sampling points are sufficient for the scalar tomography reconstruction. Indeed, the actual number of measurements is much smaller than the theoretical value. 

For a 3D vector field, all three magnetization components need to be recovered, giving a question about the particular requirement in the measured data. In his discussion in the 1980s, Norton showed that the reconstruction of a diverge-less 2D vector field appeared to be unique \cite{Norton1989}. Prince gave a more generalized discussion of the reconstructions of arbitrary vector fields in the 1990s. He demonstrated that, for reconstructing an arbitrary $n$-dimensional vector field, $n$ tomographic projection datasets in which the probe is sensitive to n different directions of the vector field needed to be acquired \cite{Prince1993, Prince1994}. The idea of using more than one tilt rotation axes has been used successfully in scalar tomography to reduce the missing wedge artifacts\cite{PENCZEK1995, ARSLAN2006}. That idea is also believed to be a key to solving the vector tomography problem.

In our research, we use the Fourier slice theorem to show specific experimental conditions for the reconstruction of arbitrary vector fields. The theorem states that the 2D Fourier transform of a 2D projection equals a 2D slice through the origin of the 3D Fourier transform of an object. The 2D slice is defined based on the corresponding rotation angle. In the case of noise-free, we apply the Fourier transform to both sides of Eqn.\ref{eqn_noise_final}. 
\begin{equation}
    \alpha_{\thetaB} \, \mathcal{F} \big[ \Pi_{\thetaB} (M_x) \big] + \beta_{\thetaB} \, \mathcal{F} \big[ \Pi_{\thetaB} (M_y) \big] + \gamma_{\thetaB} \, \mathcal{F} \big[ \Pi_{\thetaB} (M_z) \big]  = \mathcal{F} \big[  b_{\thetaB}^- \big]    
\end{equation}
Applying the Fourier slice theorem, we have a linear constraint involving the Fourier transforms $\hat{m}_x$, $\hat{m}_y$ and $\hat{m}_z$ of the three magnetization components $M_x$, $M_y$ and $M_z$.
This constraint applies to every Fourier point $\xiB$ on a 2D Fourier slice through the origin.
\begin{equation}\label{lin_eqn_Fourier}
    \alpha_{\thetaB} \, \hat{m}_x ( \xiB ) + \beta_{\thetaB} \, \hat{m}_y ( \xiB )  + \gamma_{\thetaB} \, \hat{m}_z ( \xiB ) =  \hat{b}_{\thetaB}^- (\xiB ) \quad \text{where} \quad  \langle \xiB , \nB_{\thetaB}  \rangle = 0 \text{, and } \nB_{\thetaB}  = [\alpha_{\thetaB}, \; \beta_{\thetaB}, \; \gamma_{\thetaB} ]
\end{equation}
The extra constraint $\langle \xiB , \nB_{\thetaB} ) = 0$ is required by the requirement that a point belongs to a plan through the origin if the inner product between $\xiB$ and the normal vector of the plan is zero.
If sufficient measurements are provided, one can sample the values of all Fourier points on the frequency domain and we can extend Eqn. \ref{lin_eqn_Fourier} to every point $\xiB$ on the 3D Fourier domain.

In order to separate $\hat{m}_x(\xiB)$, $\hat{m}_y(\xiB)$ and $\hat{m}_z(\xiB)$ for a given 3D frequency point $\overrightarrow{\xi}$, we need to find three 2D Fourier slices whose normal vectors $\nB_{\thetaB}$ form a linear independent system in $\mathbb{R}^3$ and that go through the origin and contains $\overrightarrow{\xi}$. This is impossible since the set of normal vectors $\nB_{\thetaB}$ that satisfies the constraint $\langle \xiB , \nB_{\thetaB}  \rangle = 0$ lies in a linear subspace of dimension two.

We give an example of in-plane rotations where the orientation is given by  $Z_{\phi}Y_{\theta}$. 
Recalling that the normal vector corresponding to the in-plane rotation has the form $\nB_{\thetaB} = (\sin \theta \, \cos \phi, \; \sin \theta \, \sin \phi, \; \cos \theta)$,
we can find infinitely many 2D slices that contain the point $\xiB =(1, \; 1,\; 1)$. For example, we name three projections with the corresponding Euler angles $(\phi_1, \; \theta_1) = (0^o, \; \shortminus 45^o)$, $(\phi_2, \; \theta_2) = (120^o, \; \shortminus 69.90^o)$ and $(\phi_3, \; \theta_3) = (-120^o, \; 36.21^o)$. 
\begin{align}    
\begin{cases}\label{lin_eqn_Fourier2}
    &\sin\theta_1 \, \cos\phi_1 \, \hat{m}_x(\xiB) + \sin \theta_1 \, \sin \phi_1 \, \hat{m}_y(\xiB)  + \cos \theta_1 \, \hat{m}_z(\xiB)  = \hat{b}_{\thetaB_1}^-(\xiB)   \\
    &\sin\theta_2 \, \cos\phi_2 \, \hat{m}_x(\xiB) + \sin \theta_2 \, \sin \phi_2 \, \hat{m}_y(\xiB)  + \cos \theta_2 \, \hat{m}_z(\xiB)  = \hat{b}_{\thetaB_2}^-(\xiB) \\
    &\sin\theta_3 \, \cos\phi_3 \, \hat{m}_x(\xiB) + \sin \theta_3 \, \sin \phi_3 \, \hat{m}_y(\xiB)  + \cos \theta_3 \, \hat{m}_z(\xiB)  = \hat{b}_{\thetaB_3}^-(\xiB) 
\end{cases}
\end{align}
One can check that the corresponding normal vectors $( \shortminus1/\sqrt{2}, \; 0, \; 1/\sqrt{2})$, (0.4695, -0.8133, 0.3437), and (-0.2953, -0.5116, 0.8069) are linearly dependent with rank two. Consequently, Eqn. \ref{lin_eqn_Fourier2} does not have a unique solution. It verifies that in-plane rotations are not sufficient for the reconstruction of the magnetization $\mB$.

This analysis differs from Norton\cite{Norton1989} and Phatak's theoretical development\cite{PHATAK2008}, which analyze the reconstruction of the magnetic vector field instead. In that case, the authors can find a linearly independent system of three equations to separate the frequency signals of the magnetic vector field $\bB$. While the first two constraints are obtained from rotations, the last constraint is found by Gauss's law $\nabla \cdot \bB = 0$ (since $\bB$ is divergence-free). 
The magnetization vector field is not divergence-free but has another important property: the magnetization field can only exist in a magnetic material.
Hence, one can utilize a support (defined as a 3D boundary of the magnetic material) as the necessary and complimentary constraint for the completeness of a magnetization reconstruction algorithm.

Furthermore, for the case of micro-magnetic and no external dynamics at the boundary, we can add in the boundary condition that the gradient of the magnetization is parallel to surface\cite{brown1963, Friedman1992, Newell1993ATM}, i.e.  $ \frac{\partial \mB} {\partial \nBB } = 0$.
In practice, since the support and boundaries are difficult to get computed exactly, one should not enforce the constraint rigidly but relax it as a regularizer instead. We add this regularizer to the minimization (\ref{min1}):
\begin{align}\label{min1_boundary}
    \min_{\mB} \; \varepsilon(\mB) &=  \frac{1}{2}\sum_{\thetaB} { \big\| \alpha_{\thetaB} \, \Pi_{\thetaB} (M_x) + \beta_{\thetaB} \, \Pi_{\thetaB} (M_y) + \gamma_{\thetaB} \, \Pi_{\thetaB} (M_z)  - b_{\thetaB}^-  \big\|^2  + \frac{\epsilon}{2} \|  \nabla \mB \cdot \nBB_{\partial \Omega} \|^2_{\partial \Omega}}
\end{align}
For the regularizer part, $\nBB_{\partial \Omega} = (n_1, n_2, n_3)$ is the normal vector to the boundary surface $\partial \Omega$ of the  magnetic sample and $\epsilon$ is the regularizer parameter. 
The regularizer term $\|  \nabla \mB \cdot \nBB \|^2_{\partial \Omega}$ only takes places on the boundary and should not affect the magnetization within the magnetic structure. 
In further expansion, we can write the regularizer explicitly as 
$  \|  \nabla \mB \cdot \nBB_{\partial \Omega} \|^2_{\partial \Omega} = \big\| n_1\frac{\partial M_x}{\partial x}  + n_2 \frac{\partial M_y}{\partial y}  + n_3 \frac{\partial M_z}{\partial z} \big\|^2_{\partial \Omega}$. 
$\epsilon$ is tunable and should be small for non-exact support. We can even ignore this regularizer (or set $\epsilon=0$) when the support cannot be computed accurately. In contrast, we can choose large $\epsilon$ for larger effect if the exact support is given. The final gradient will get computed with the extra term as below:
\begin{align}\label{grad_mx_boundary}
    \frac{\partial \varepsilon}{\partial M_x} 
     = \sum_{\thetaB} \alpha_{\thetaB} \, \Pi_{\thetaB}^T \Big( \Pi_{\thetaB} \big( \alpha_{\thetaB} M_x + \beta_{\thetaB}  \, M_y + \gamma_{\thetaB}  \, M_z \big)  - b_{\thetaB}^- \Big)  + \epsilon \; n_1 \; \frac{\partial ^T}{\partial x} \Big( n_1\frac{\partial M_x}{\partial x}  + n_2  \frac{\partial M_y}{\partial y}  + n_3 \frac{\partial M_z}{\partial z}  
 \Big)_{\partial \Omega}
\end{align}

Next, we discuss the robustness of each magnetization component reconstruction $M_x$, $M_y$, and $M_z$ concerning the in-plane rotation angles $\phi$.
The linear constraint for in-plane rotations as in Eqn. \ref{lin_eqn_Fourier2} reveals that the x and y components are coupled by linear factors $\sin\theta \, \cos\phi$ and $\sin\theta \, \sin\phi$ while the linear factor of z is only $\cos\theta$. So the x and y parts are coupled at a higher degree than the z component. As a result, the z component will get decoupling easier and yield a high-quality reconstruction than the other two. 
Assuming that two in-plane rotation $\phi_1$ and $\phi_2$ are chose, then these two angles should be chosen equally distanced on half of the unit circle to improve the robustness of the reconstruction. We can choose $\phi_1=0^o$ and $\phi_2=90^o$ as a simple option.

Side rotations can improve the robustness of the x and y components, but that approach is experimentally infeasible. The summary of our analysis is shown below:
\begin{enumerate}
    \item In-plane rotations are necessary but not sufficient to decouple the Fourier coefficients of the three magnetization components.
    \item Other constraints, such as support and boundary constraints, and regularizers, need invoking, if possible, for highly accurate reconstruction.
    \item The z component gets reconstructed with higher quality than the x and y components in in-plane rotation systems.
\end{enumerate}
With the help of a support constraint, we will show that highly accurate vector tomography reconstruction can be obtained numerically with in-plane rotations.

\section*{Vector tomography reconstruction of simulated data}
In this simulation, the sample is a meta-lattice with a size of $100 \times 100 \times 100$ pixels. The signals of the magnetization make up around $1.65\%$ of the total signal. Two tilt series from two in-plane rotations where $\phi=0^o$ and $90^o$ are inspected. For each tilt series, 45 projections of each left and right polarization $P_{\thetaB}^+$ and $P_{\thetaB}^-$ are generated in the range of $[-66^o,66^o]$ with an increment of 3 degrees. So totally, we generate 180 projections with size $100 \times 100$ pixels. To make it realistic, we add Poisson noise to the projections by selecting a flux of $4e8$ photons. This flux yields an SNR of 200 and less than $1\%$ noise. Reconstructing the non-magnetic part is not the interest of this research. However, we need to assume that the support of the non-magnetic part is given since it plays an essential role in the reconstruction of the magnetization $\mB$.
\begin{figure}
    \centering
    \begin{tabular}{cccc}
         & $M_x$ & $M_y$ & $M_z$ \\
     \rotatebox{90}{\hspace{0.8cm}model} &
    
    \begin{tikzpicture}
    \draw (0, 0) node[inner sep=0] {\includegraphics[width=2.8cm]{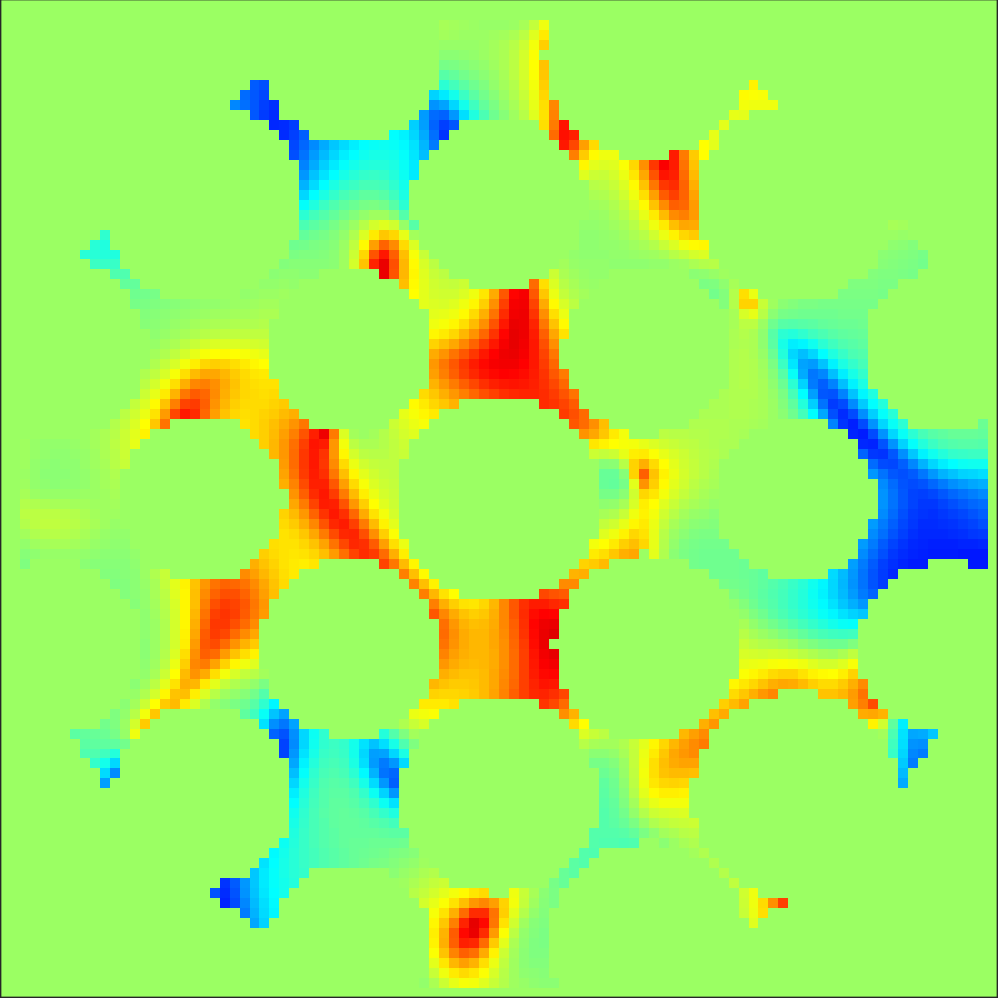}};
    \draw (-1.2, 1.1)node{\textcolor{black}{\textbf{ a}}};
    \end{tikzpicture}&

    \begin{tikzpicture}
    \draw (0, 0) node[inner sep=0] {\includegraphics[width=2.8cm]{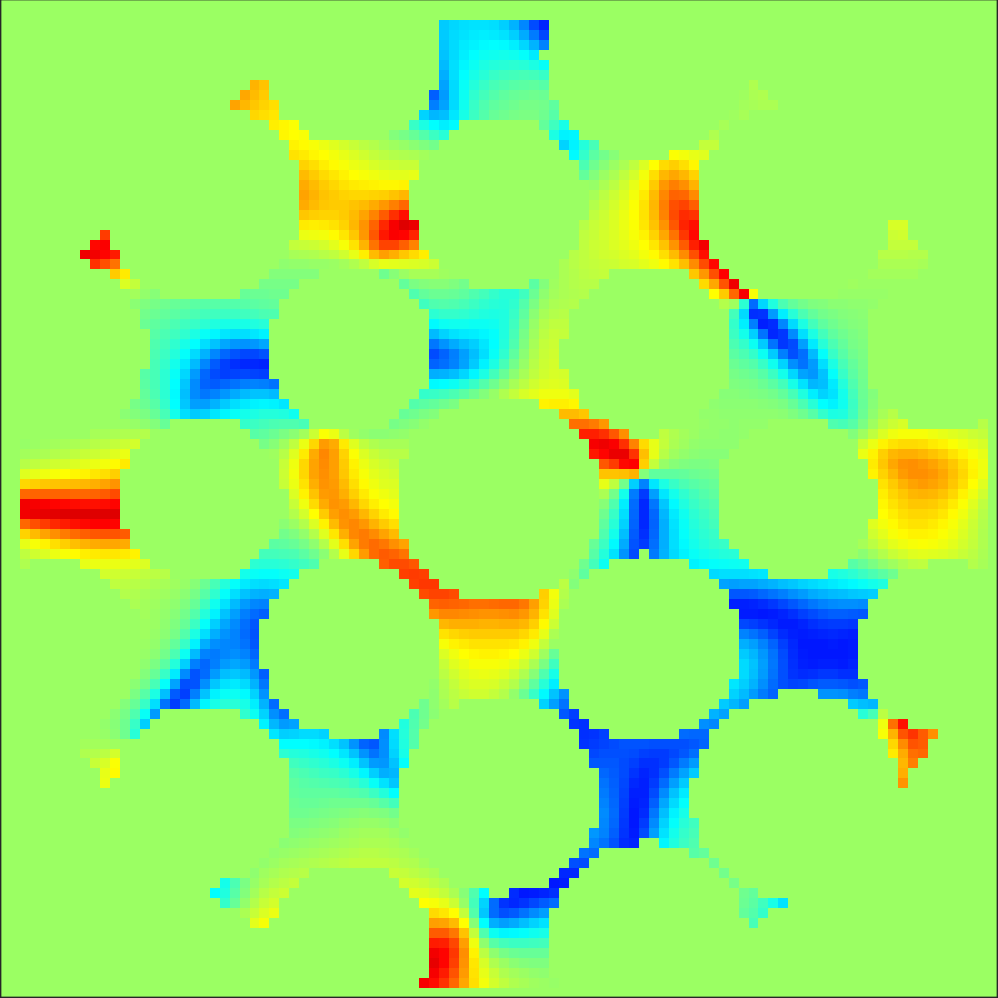}};
    \draw (-1.2, 1.1)node{\textcolor{black}{\textbf{ b}}};
    \end{tikzpicture}&

    \begin{tikzpicture}
    \draw (0, 0) node[inner sep=0] {\includegraphics[width=2.8cm]{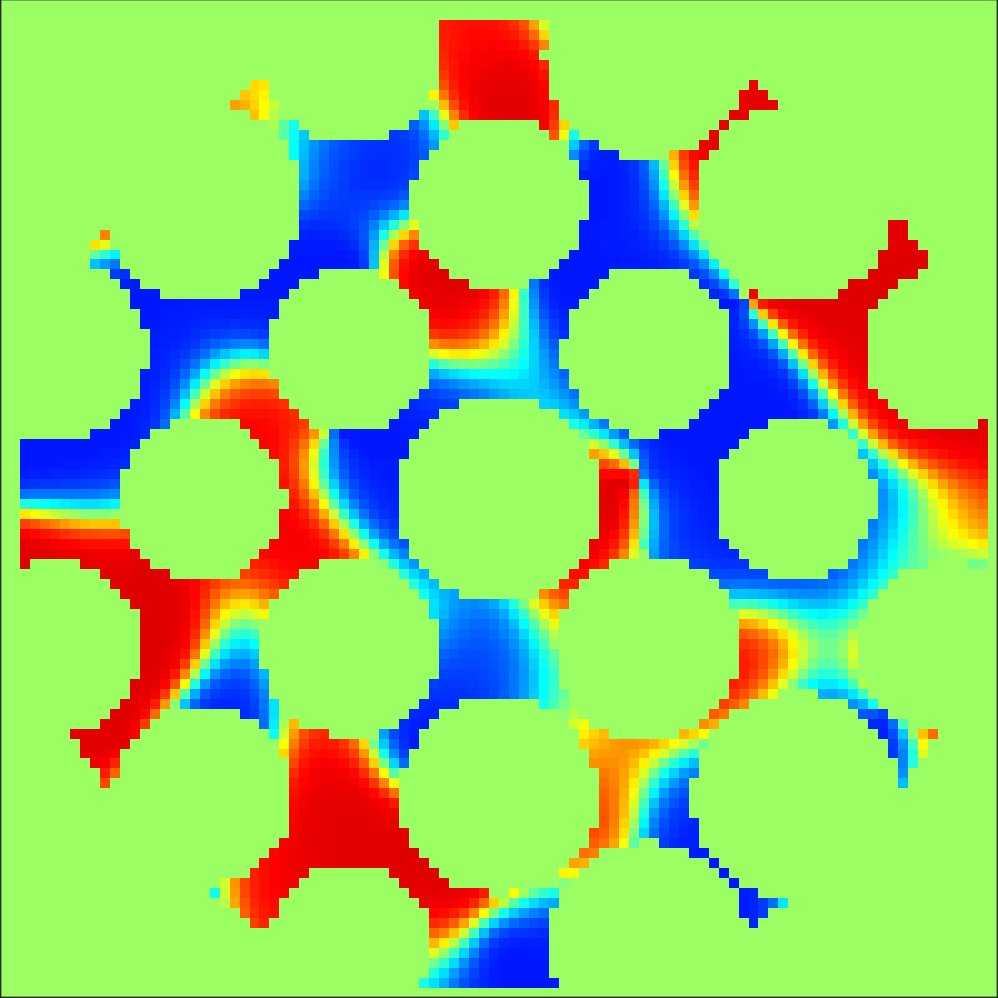}};
    \draw (-1.2, 1.1)node{\textcolor{black}{\textbf{ c}}};
    \end{tikzpicture} \\

    \rotatebox{90}{\hspace{0.3cm}reconstruction} &

    \begin{tikzpicture}
    \draw (0, 0) node[inner sep=0] {\includegraphics[width=2.8cm]{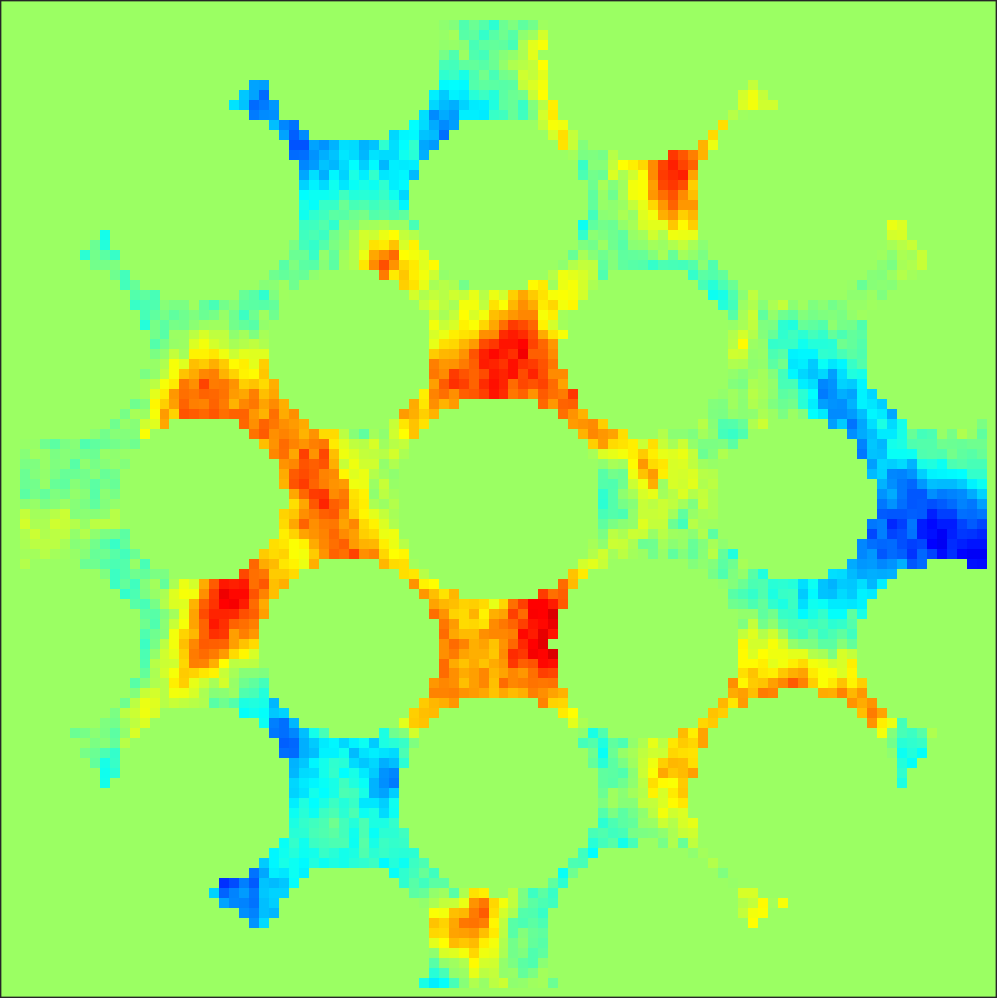}};
    \draw (-1.2, 1.1)node{\textcolor{black}{\textbf{ d}}};
    \end{tikzpicture}&

    \begin{tikzpicture}
    \draw (0, 0) node[inner sep=0] {\includegraphics[width=2.8cm]{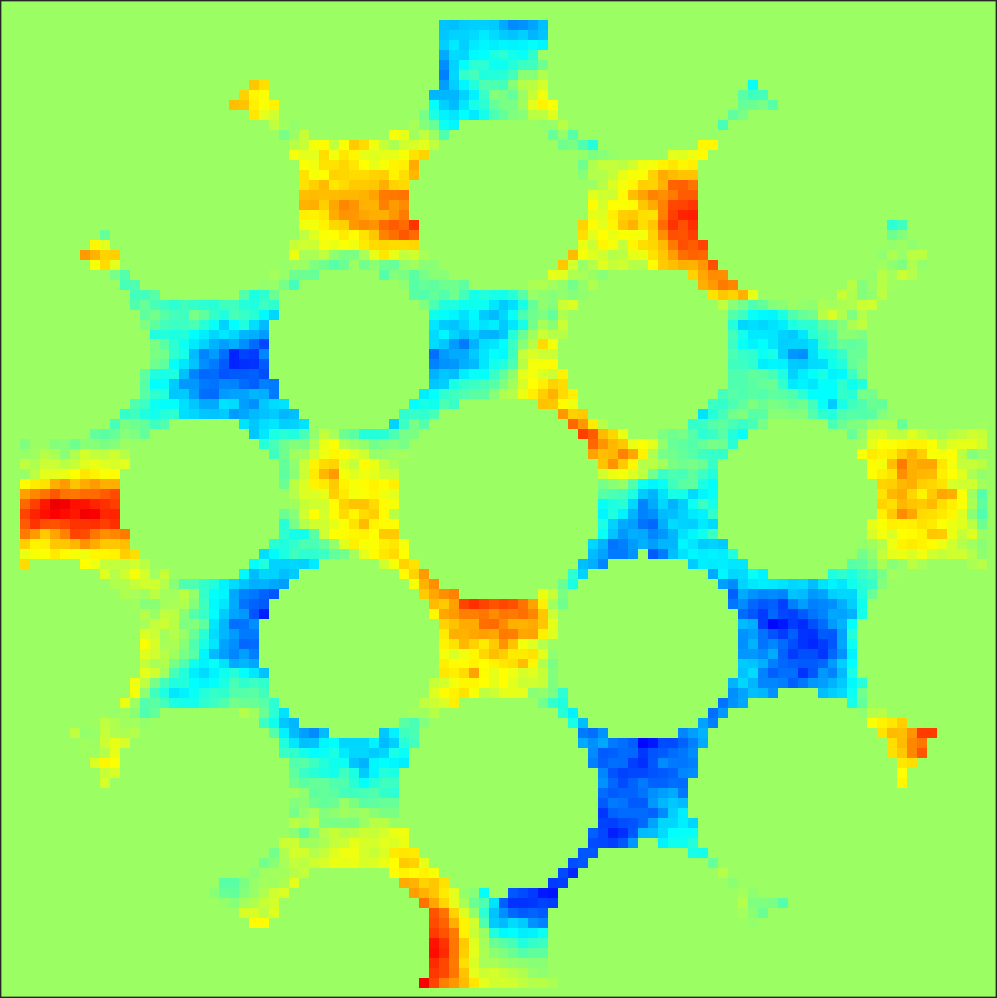}};
    \draw (-1.2, 1.1)node{\textcolor{black}{\textbf{ e}}};
    \end{tikzpicture}&

    \begin{tikzpicture}
    \draw (0, 0) node[inner sep=0] {\includegraphics[width=2.8cm]{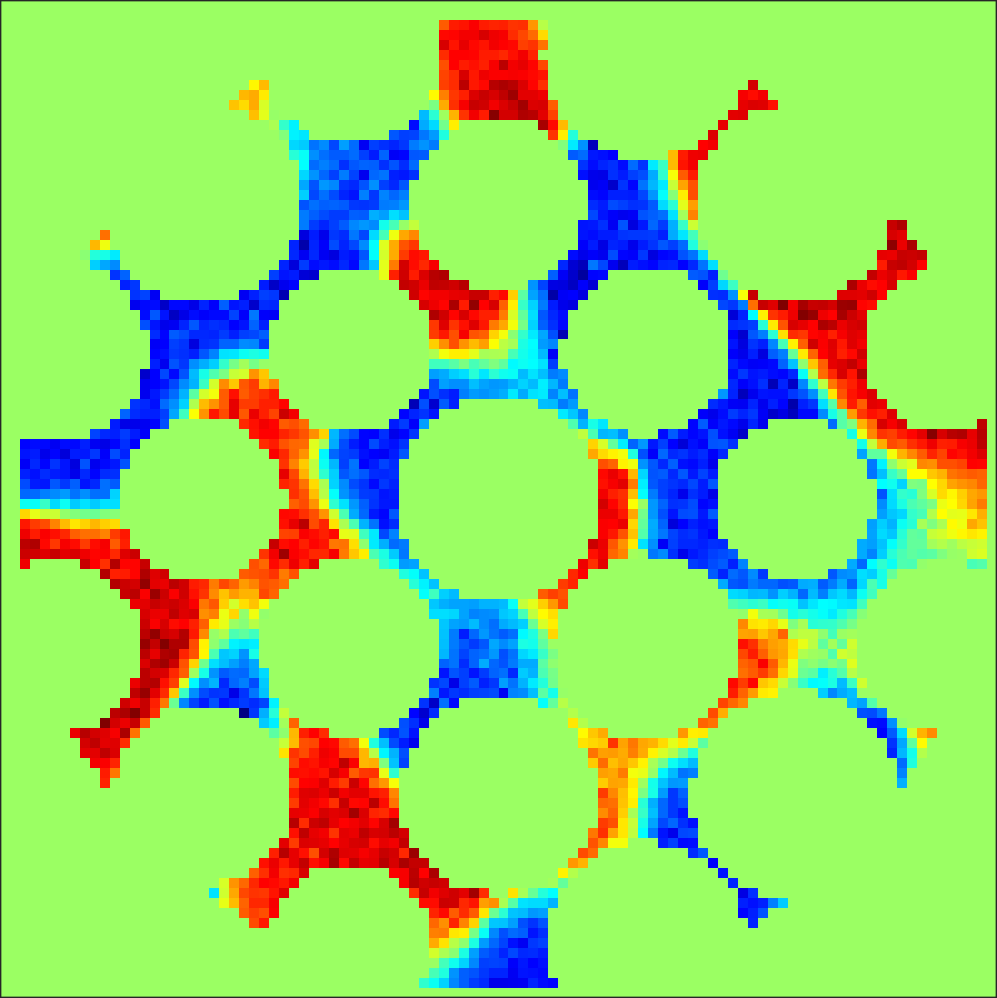}};
    \draw (-1.2, 1.1)node{\textcolor{black}{\textbf{ f}}};
    \end{tikzpicture}
    \end{tabular}
    \raisebox{-2.7cm}{\begin{tikzpicture}
    \draw (0, 0) node[inner sep=0] {\includegraphics[height=5.4cm]{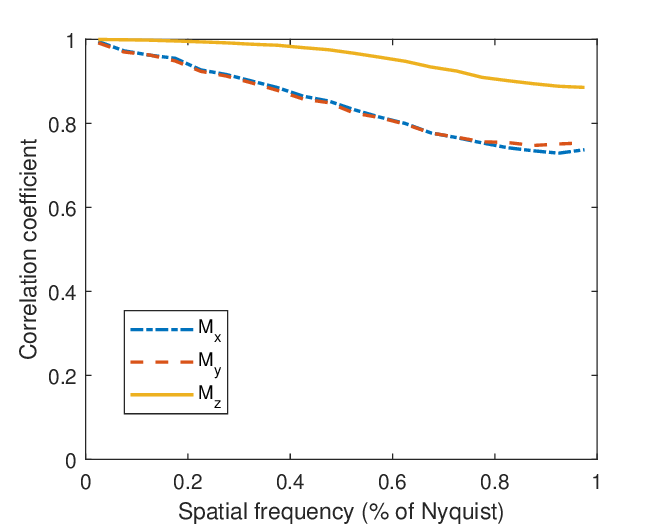}};
    \draw (-2.2, 1.8)node{\textcolor{black}{\textbf{g}}}; 
    \end{tikzpicture}}\\
    \begin{tikzpicture}
    \draw (0, 0) node[inner sep=0] {\includegraphics[width=3.5cm]{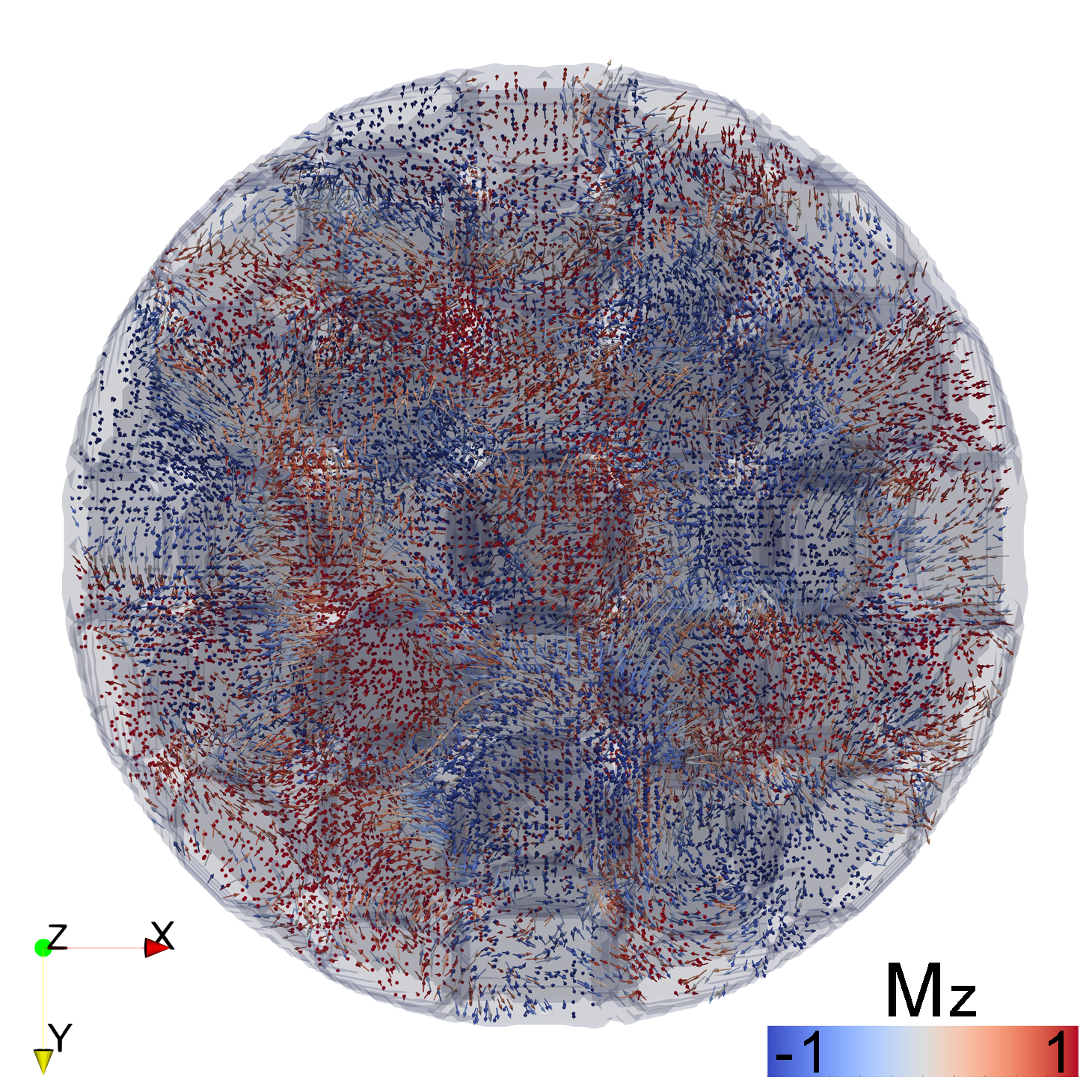}};
    \draw (-1.4, 1.4)node{\textcolor{black}{\textbf{ h}}}; 
    \end{tikzpicture}
    \begin{tikzpicture}
    \draw (0, 0) node[inner sep=0] {\includegraphics[width=3.5cm]{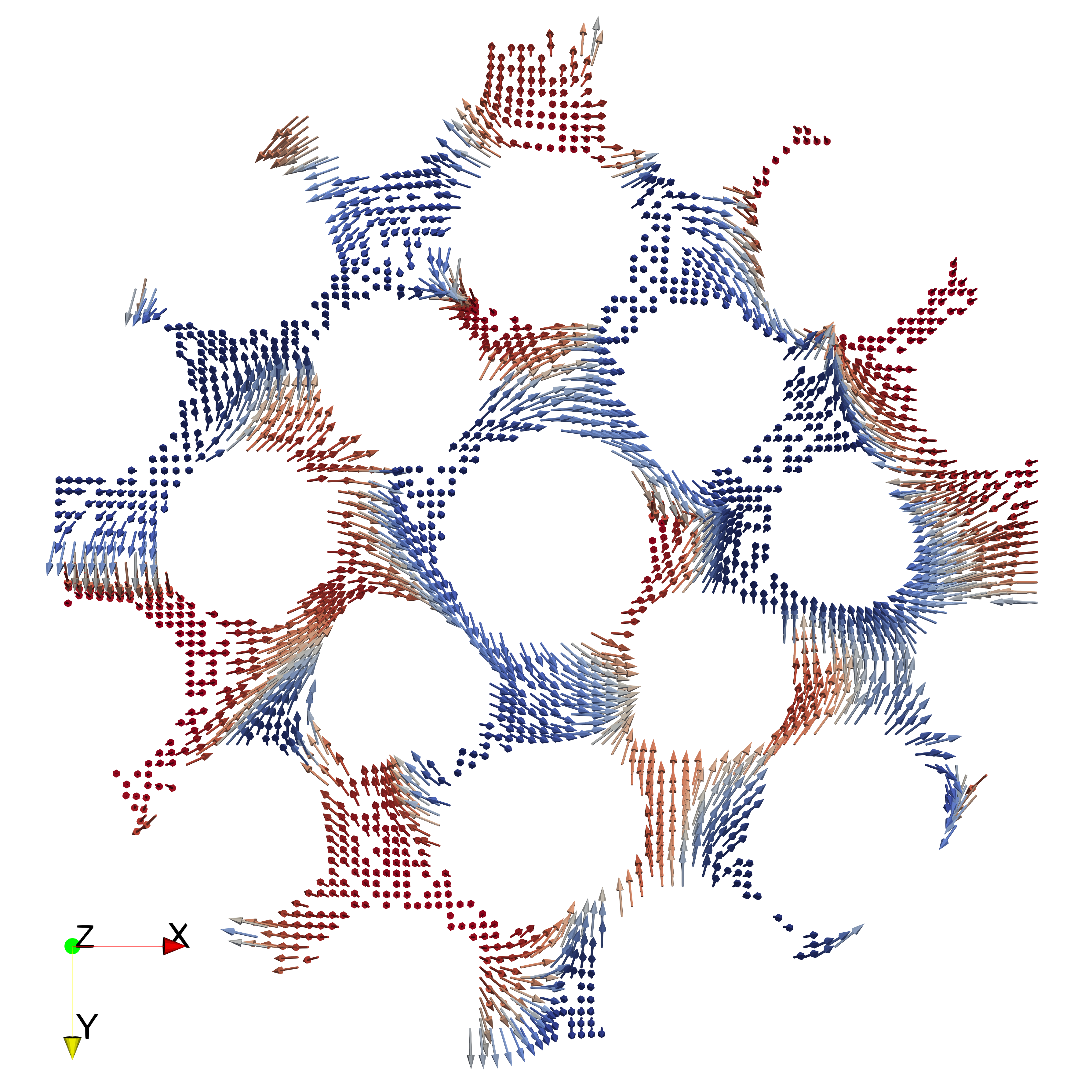}};
    \draw (-1.4, 1.4)node{\textcolor{black}{\textbf{ i}}};
    \end{tikzpicture} 
    \begin{tikzpicture}
    \draw (0, 0) node[inner sep=0] {\includegraphics[width=3.5cm]{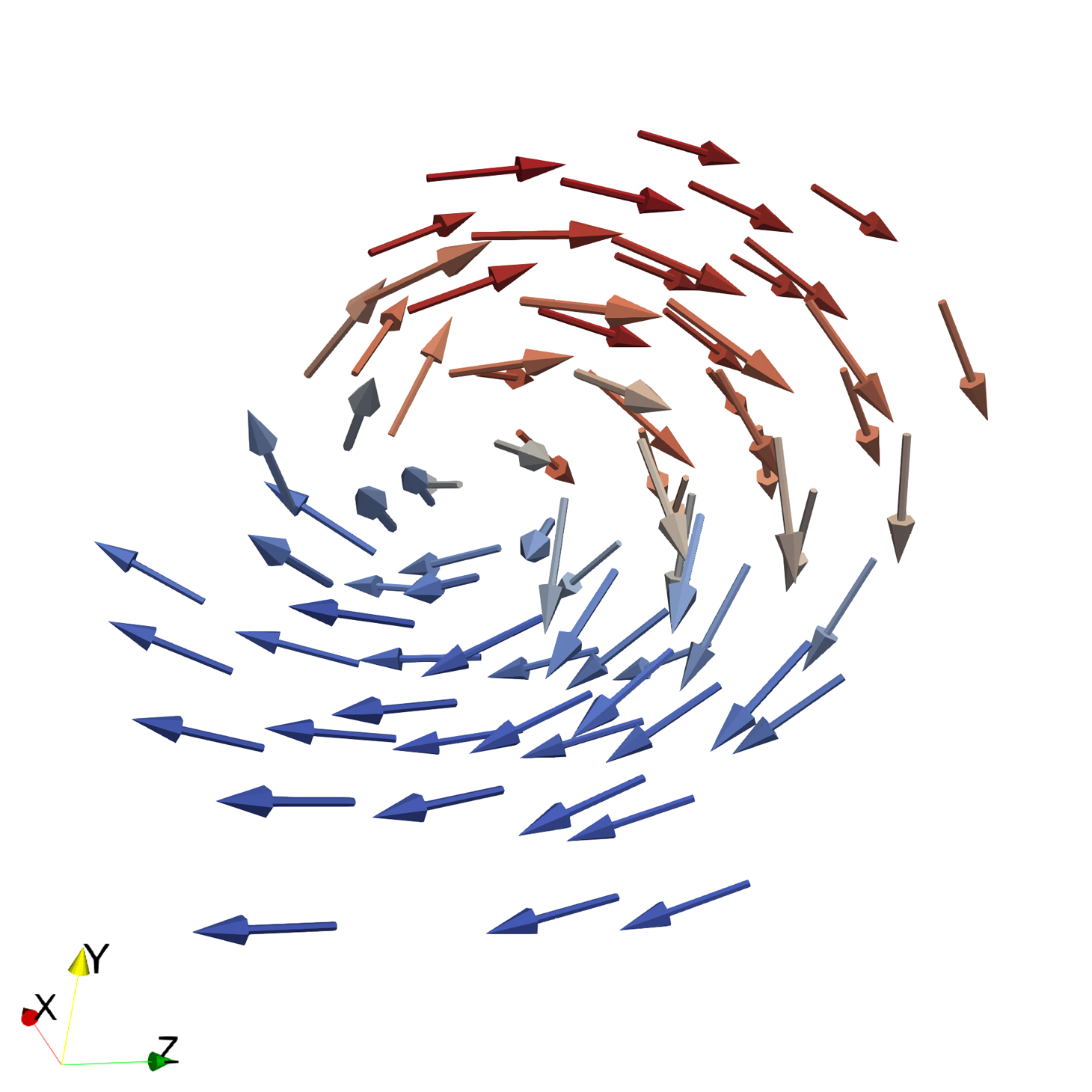}};
    \draw (-1.4, 1.4)node{\textcolor{black}{\textbf{ j}}};
    \end{tikzpicture}
    \begin{tikzpicture}
    \draw (0, 0) node[inner sep=0] {\includegraphics[width=3.5cm]{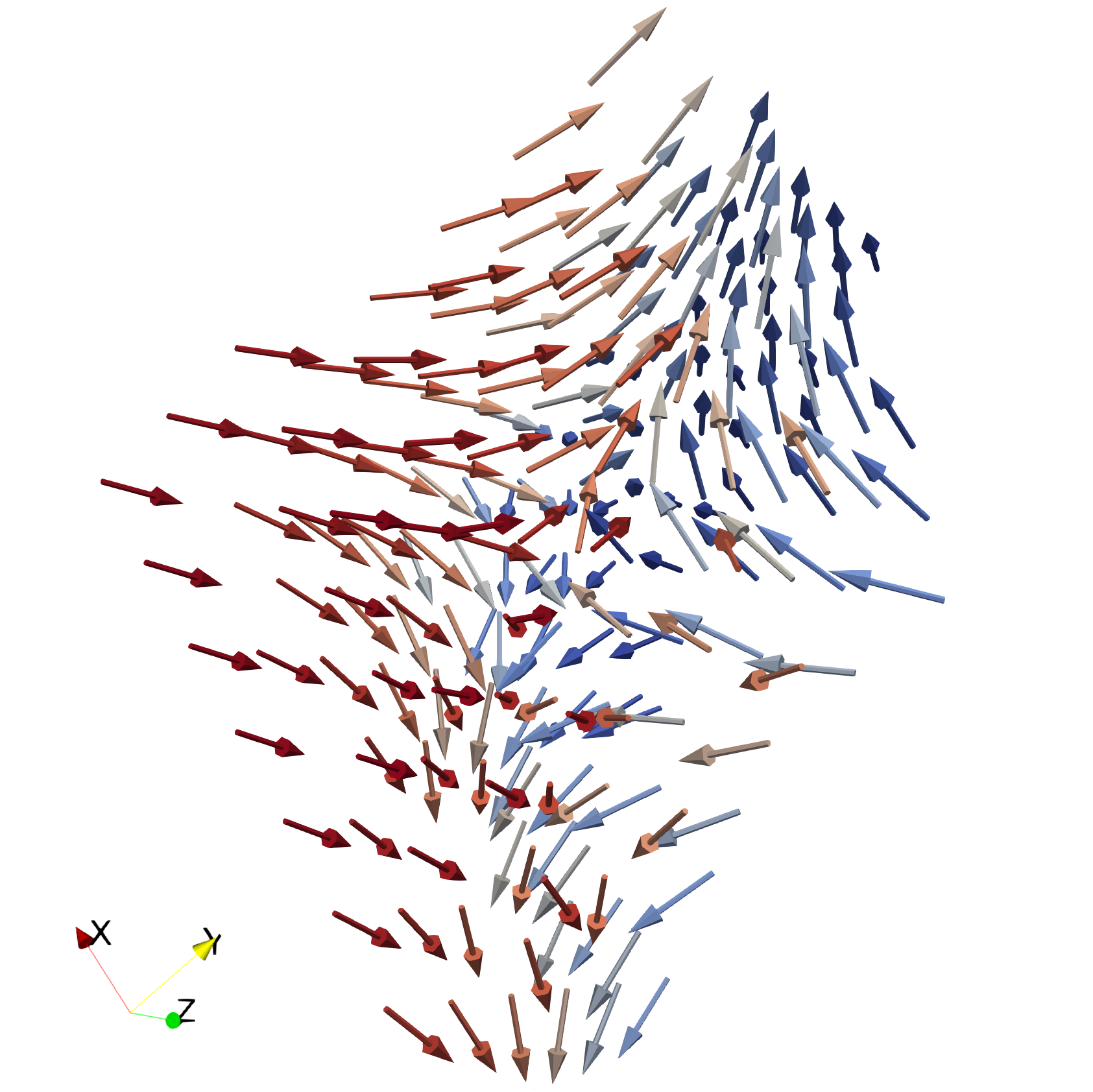}};
    \draw (-1.4, 1.4)node{\textcolor{black}{\textbf{ k}}};
    \end{tikzpicture}\\

    \begin{tikzpicture}
    \draw (0, 0) node[inner sep=0] {\includegraphics[width=3.5cm]{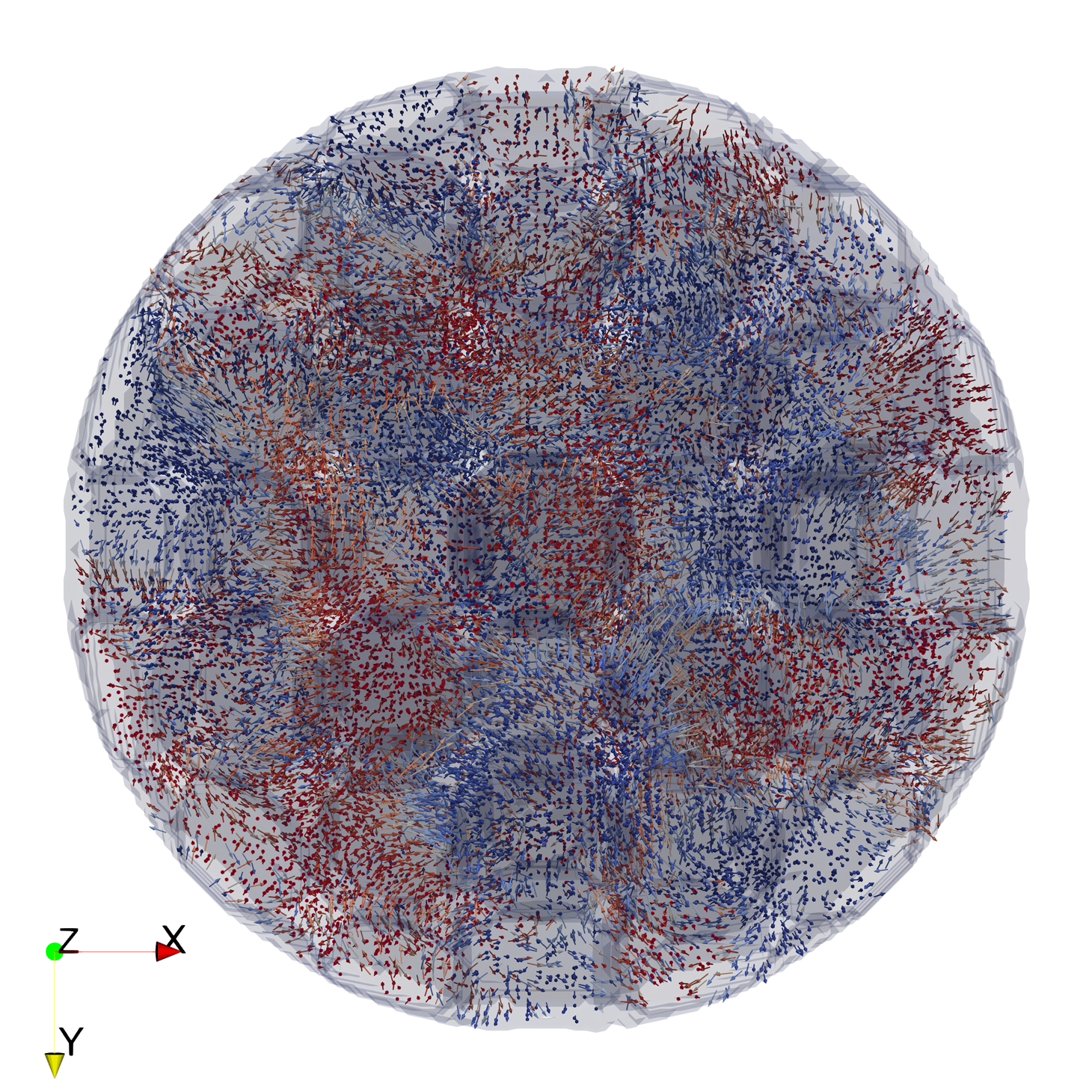}};
    2\draw (-1.4, 1.4)node{\textcolor{black}{\textbf{l}}};
    \end{tikzpicture}
    \begin{tikzpicture}
    \draw (0, 0) node[inner sep=0] {\includegraphics[width=3.5cm]{images/slice51_model.png}};
    \draw (-1.4, 1.4)node{\textcolor{black}{\textbf{m}}};
    \end{tikzpicture} 
    \begin{tikzpicture}
    \draw (0, 0) node[inner sep=0] {\includegraphics[width=3.5cm]{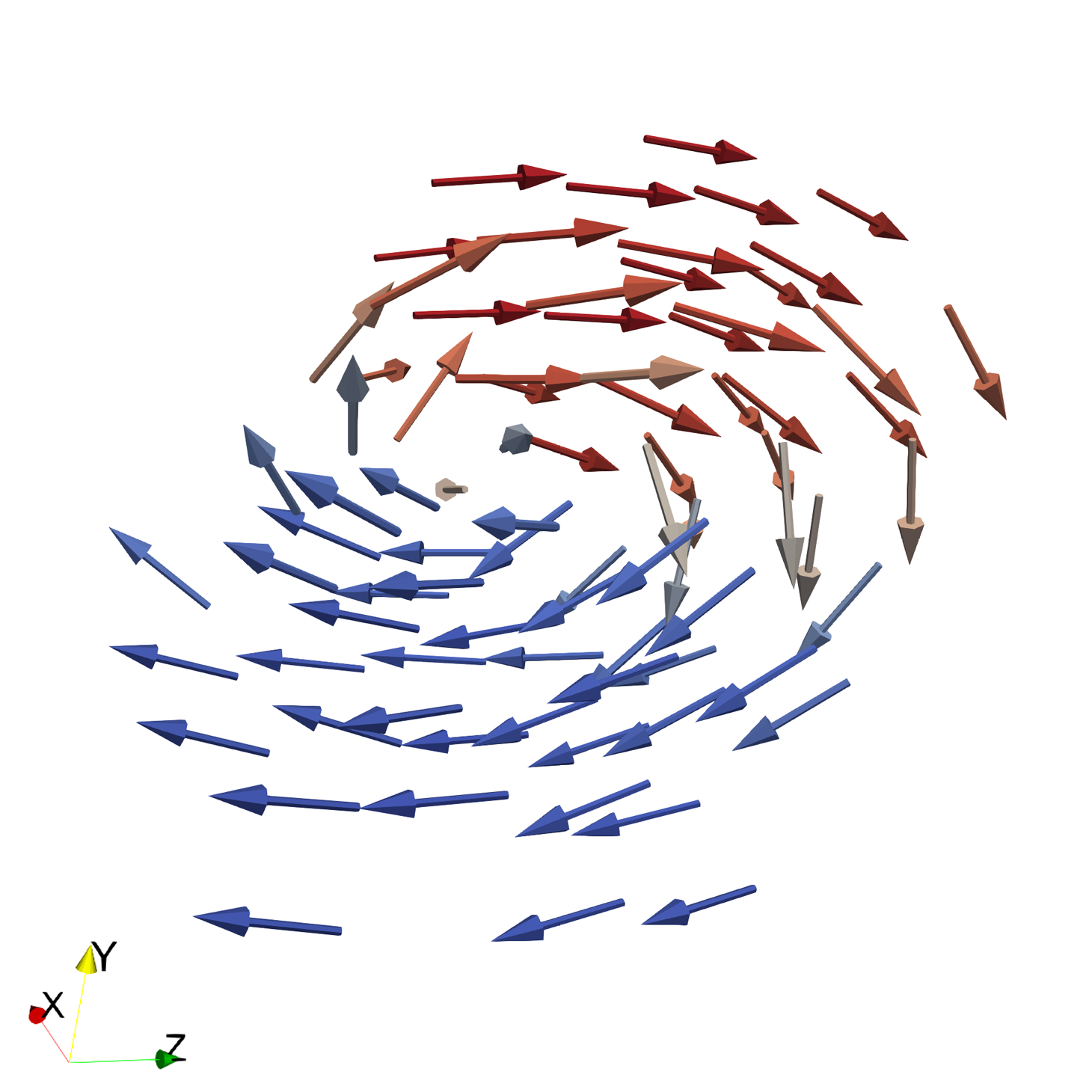}};
    \draw (-1.4, 1.)node{\textcolor{black}{\textbf{n}}};
    \end{tikzpicture}
    \begin{tikzpicture}
    \draw (0, 0) node[inner sep=0] {\includegraphics[width=3.5cm]{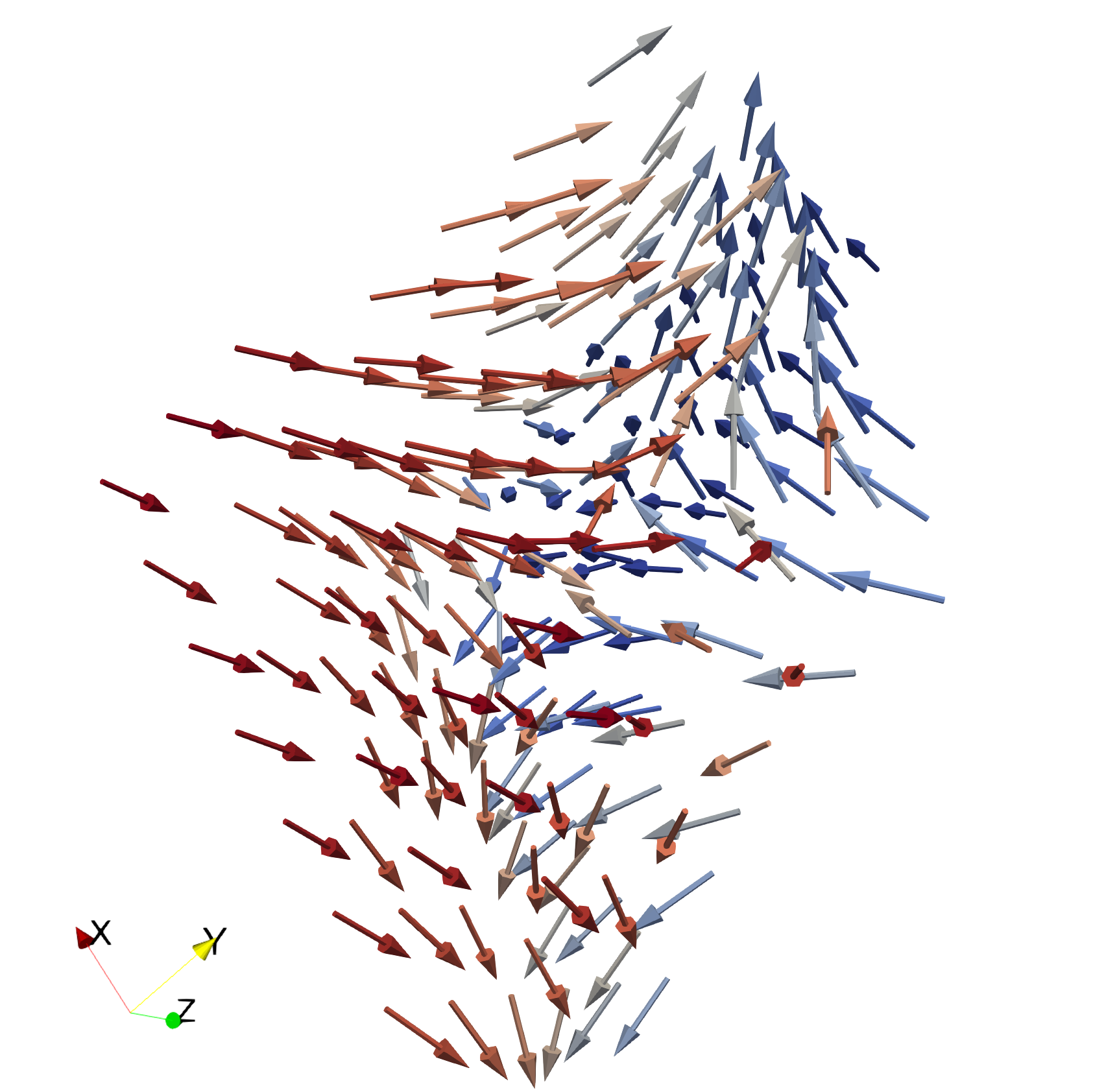}};
    \draw (-1.4, 1.4)node{\textcolor{black}{\textbf{o}}};
    \end{tikzpicture}
    
    \caption{Vector tomography reconstruction of simulated data. (\textbf{a-f}) Three magnetic components at the central slice in the z direction of the model (\textbf{a-c}) and vector tomography reconstruction (\textbf{d-f}) from the simulated data where the normalized cross correlations  are 94.1\%, 93.8\% and 99.1\% for $M_x$, $M_y$ and $M_z$ respectively. (\textbf{g}) Fourier shell correlation of the three magnetic components,  also confirming that the z component has higher quality than the x and y components. (\textbf{h-k}) 3D magnetization vector field of the model, including the overall vector field (\textbf{h}), the central slice along the z direction (\textbf{i}), two topological defect with positive charge (\textbf{j}) and negative charge (\textbf{k}), where the colors represent the different directions of the vectors. (\textbf{l-o}) Reconstructed 3D magnetization vector filed, including the overall vector field (\textbf{l}), the central slice along the z direction (\textbf{m}), two topological defects with positive charge (\textbf{n}) and negative charge (\textbf{o}), which are in good agreement with (\textbf{h-k}).
    }
    \label{fig1}
\end{figure}

Next, we take the left and right projection difference $b_{\thetaB}^- = \frac{1}{2}( P_{\thetaB}^+ - P_{\thetaB}^-)$. Since the magnetic part only makes up a fraction of the total signal, its SNR is much smaller than that of the non-magnetic part. The SNR of the projection difference is approximately $1.65\% \times 200 = 3.3$, which is quite small. The high noise level in the projection differences causes the reconstruction of the magnetization $\mB$ to be less robust than the scalar one. 
Assuming that the noise level in the non-magnetic part stays the same, then the robustness of the reconstruction will decline as the magnetization signals decrease relative to the non-magnetic signals.

Now we use our algorithm to reconstruct the three magnetization components $M_x$, $M_y$, and $M_z$. The model and result are shown in Fig. \ref{fig1}. Since the support constraint is enforced, that is, the magnetization field only appears in the magnetic material. 
In addition, since we use two tilt series at $\phi=0^o$ and $\phi=90^o$, the missing wedge artifact does not significantly affect the reconstruction. Fig. \ref{fig1}d-f, shows $M_x$, $M_y$, and $M_z$ components in the central slice along the z-axis, which are in good agreement with the model the qualities of reconstructions in all directions are comparable and as good as the model (Fig. \ref{fig1}a-c).
To quantify the vector tomography reconstruction, we calculate the Fourier shell correlation of the three components between the model and the reconstruction (Fig. \ref{fig1}g). The large correlation coefficients indicate the good quality of the vector tomography reconstruction. Additionally, we observe that the reconstructed $M_z$ has higher quality than $M_x$ and $M_y$, which is consistent with our analysis.
Fig. \ref{fig1}h-i show the 3D magnetization vector field of the reconstruction and the central slice along the z-axis, respectively, which agree with the model (Fig. \ref{fig1}l-m). 
We also plot two topological defects with one positive topological charge (Fig. \ref{fig1}j) and the other negative charge (Fig. \ref{fig1}k), both of which are in accordance with the model (Fig. \ref{fig1}n-o). 
All these analyses confirm that RESIRE-V can reconstruct a high-quality 3D vector field from multiple tilt series each with a limited number of projections with a missing wedge.

\section*{Vector tomography reconstruction of an experimental data of a ferromagnetic meta-lattice}
\begin{figure}
    \centering
    \begin{tabular}{cc}
        \begin{tikzpicture}
            \draw (0, 0) node[inner sep=0] {\includegraphics[height=2.5cm, scale=0.5]{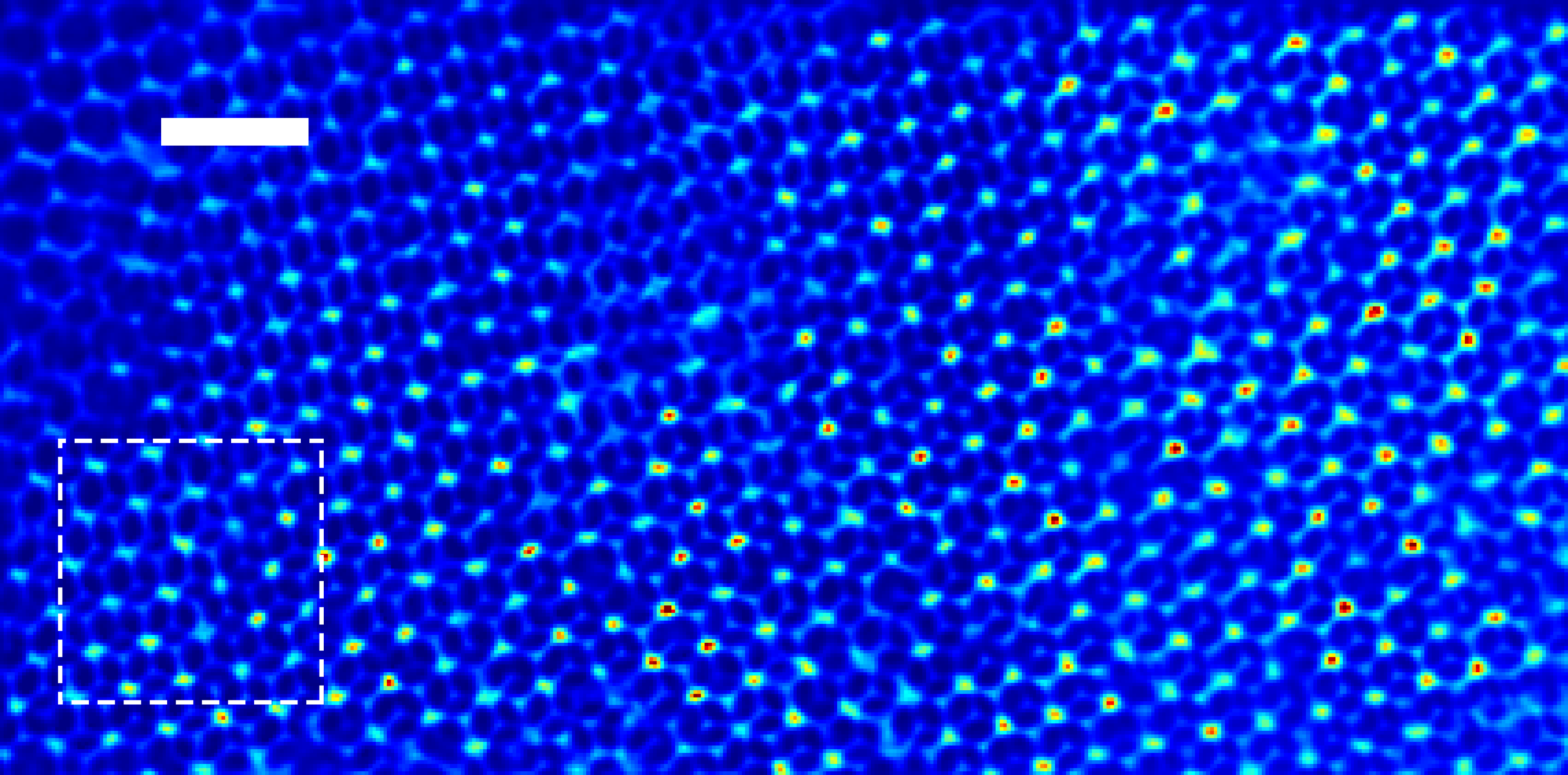}};
            \draw (-2.4, 1.0)node{\textcolor{white}{\textbf{a}}};
        \end{tikzpicture}
        \includegraphics[height=2.5cm, scale=0.5]{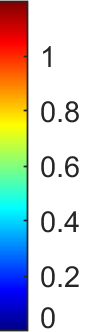}\\
         \begin{tikzpicture}
            \draw (0, 0) node[inner sep=0] {\includegraphics[height=2.5cm, scale=0.5]{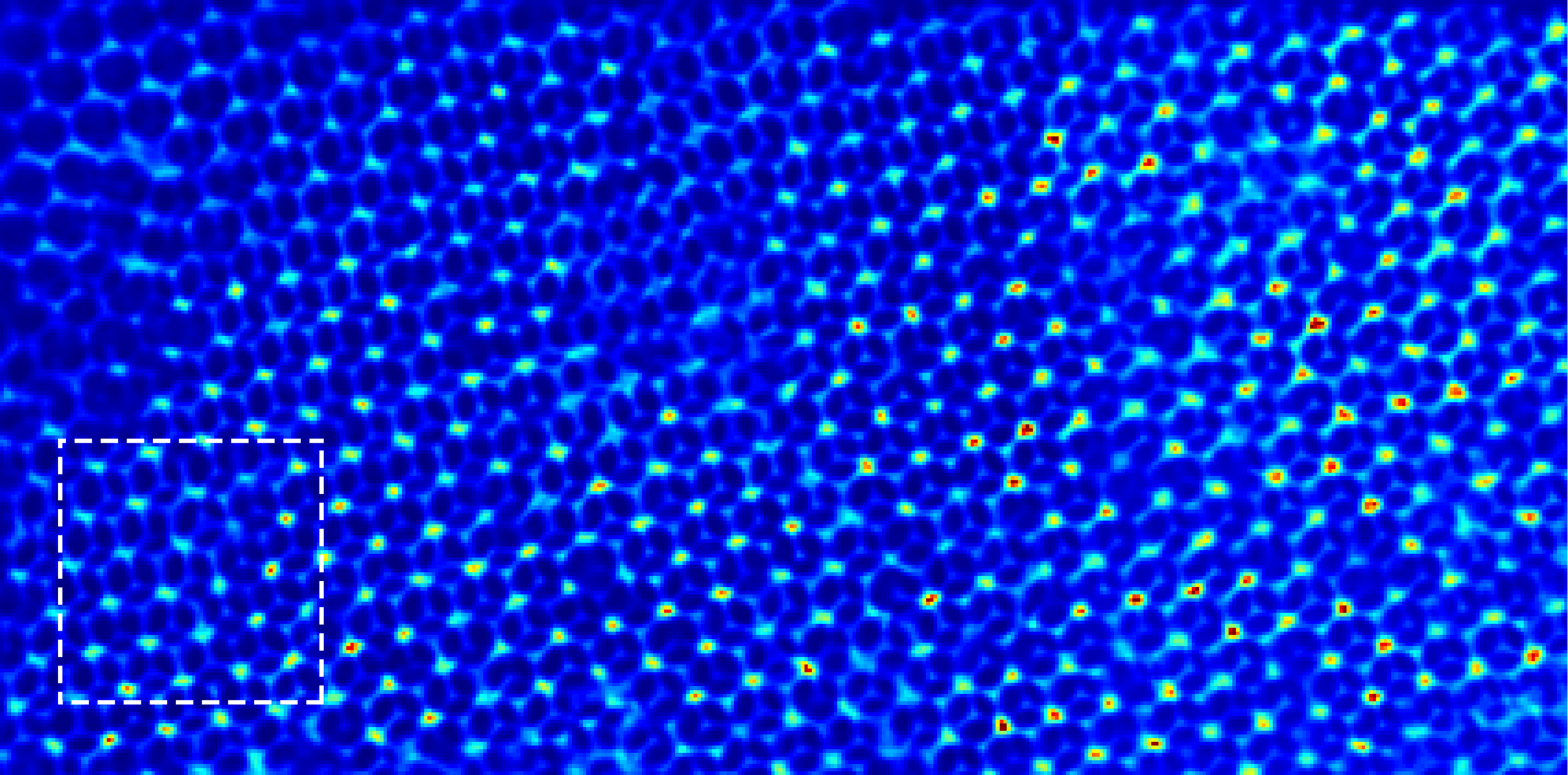}};
            \draw (-2.4, 1.0)node{\textcolor{white}{\textbf{b}}};
        \end{tikzpicture}
        \includegraphics[height=2.5cm, scale=0.5]{colorbar3a.png}\\
        \begin{tikzpicture}
            \draw (0, 0) node[inner sep=0] {\includegraphics[height=2.5cm, scale=0.5]{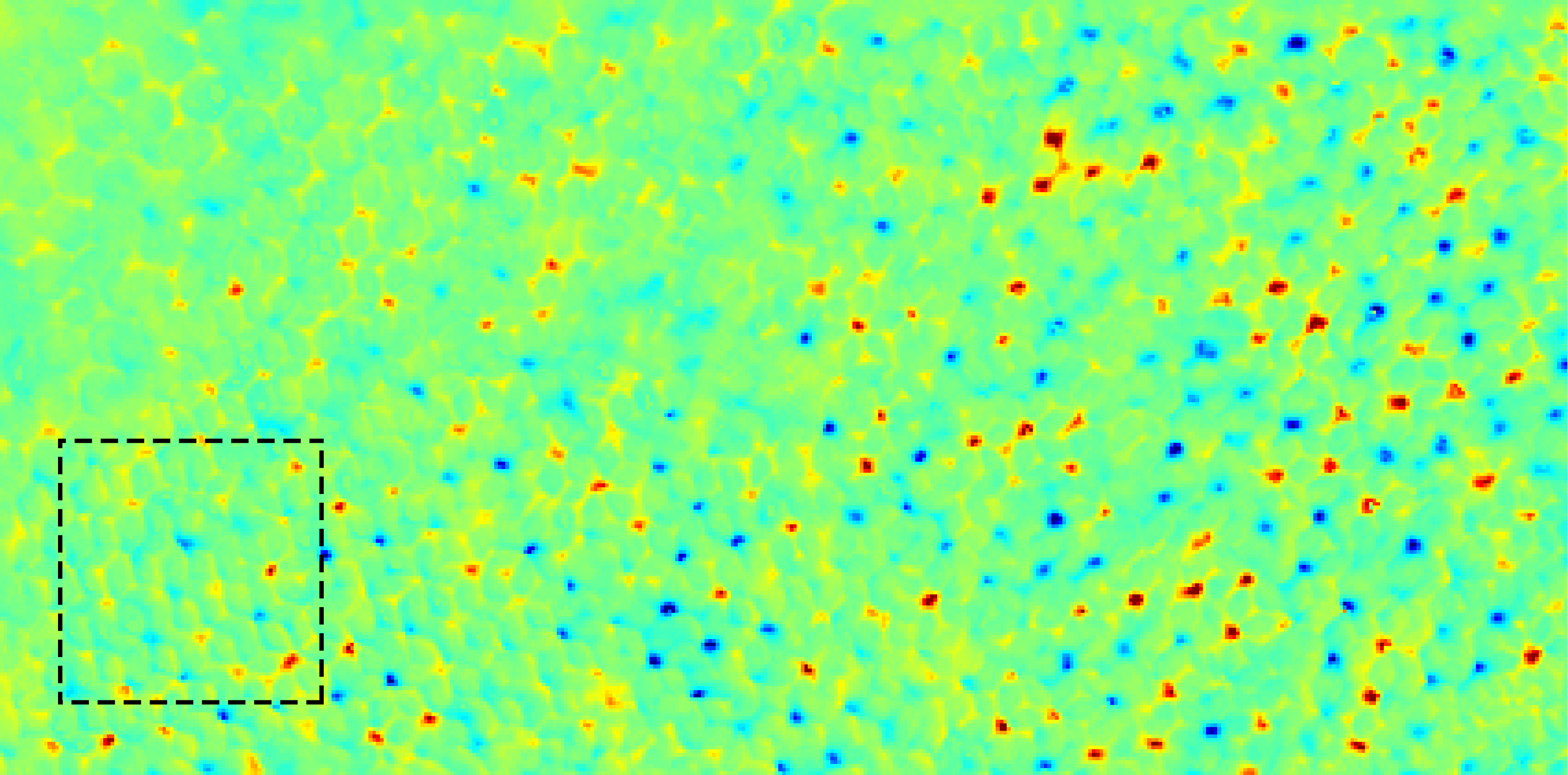}};
            \draw (-2.4, 1.0)node{\textcolor{black}{\textbf{c}}};
        \end{tikzpicture} 
        \includegraphics[height=2.5cm, scale=0.5]{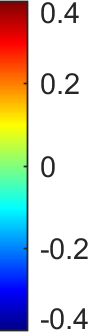}\\
    \end{tabular}
    \begin{tabular}{c}
        \begin{tikzpicture}
            \draw (0, 0) node[inner sep=0] {\includegraphics[height=4.2cm, scale=0.5]{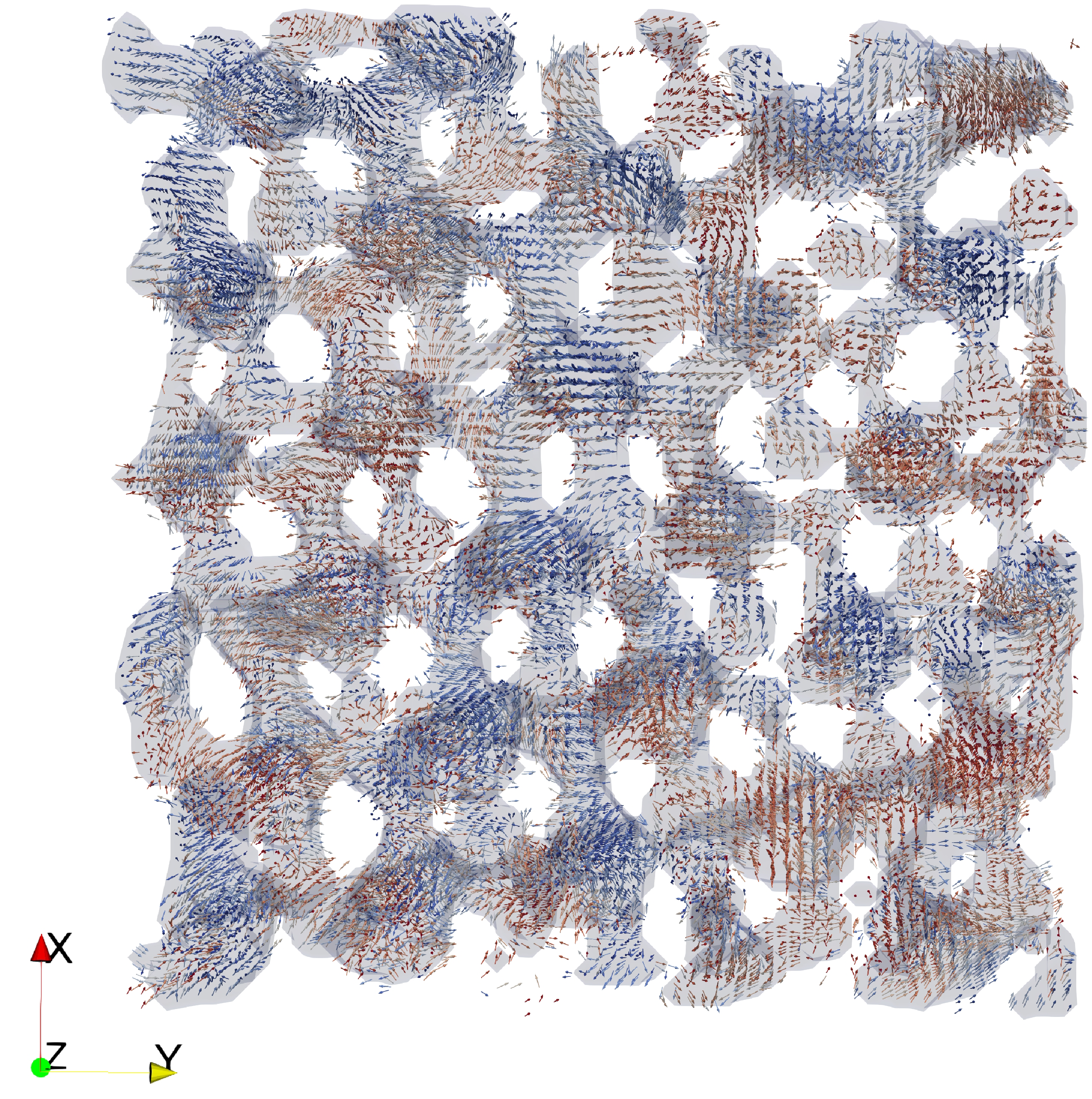}};
            \draw (-2.0, 1.6)node{\textcolor{black}{\textbf{d}}};
        \end{tikzpicture}   \\       
        \begin{tikzpicture}
            \draw (0, 0) node[inner sep=0] {\includegraphics[height=3.3cm]{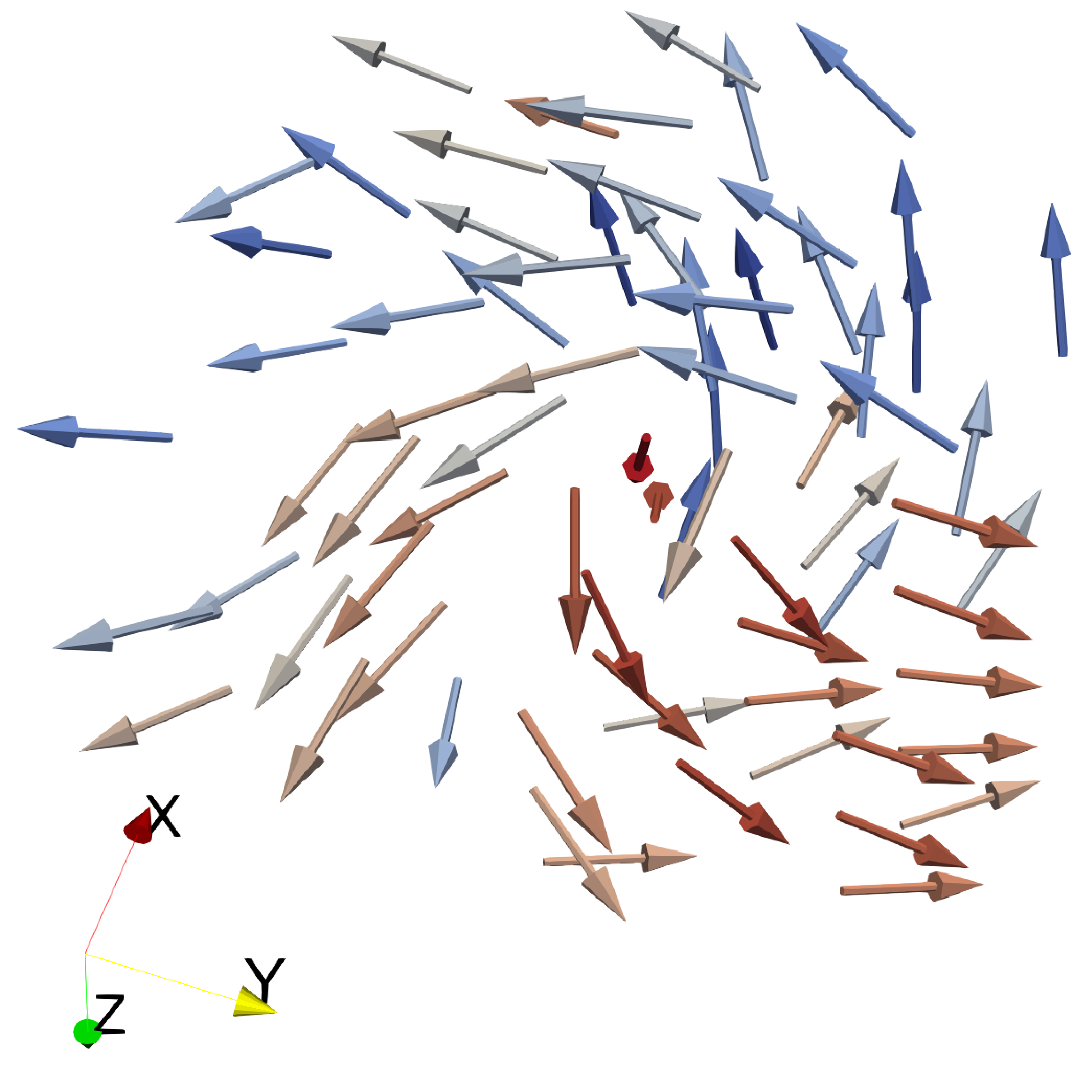}};
            \draw (-1.6, 1.4)node{\textcolor{black}{\textbf{f}}};
        \end{tikzpicture} 
    \end{tabular}
    \begin{tabular}{c}
        \begin{tikzpicture}
            \draw (0, 0) node[inner sep=0] {\includegraphics[height=5.9cm, scale=0.5]{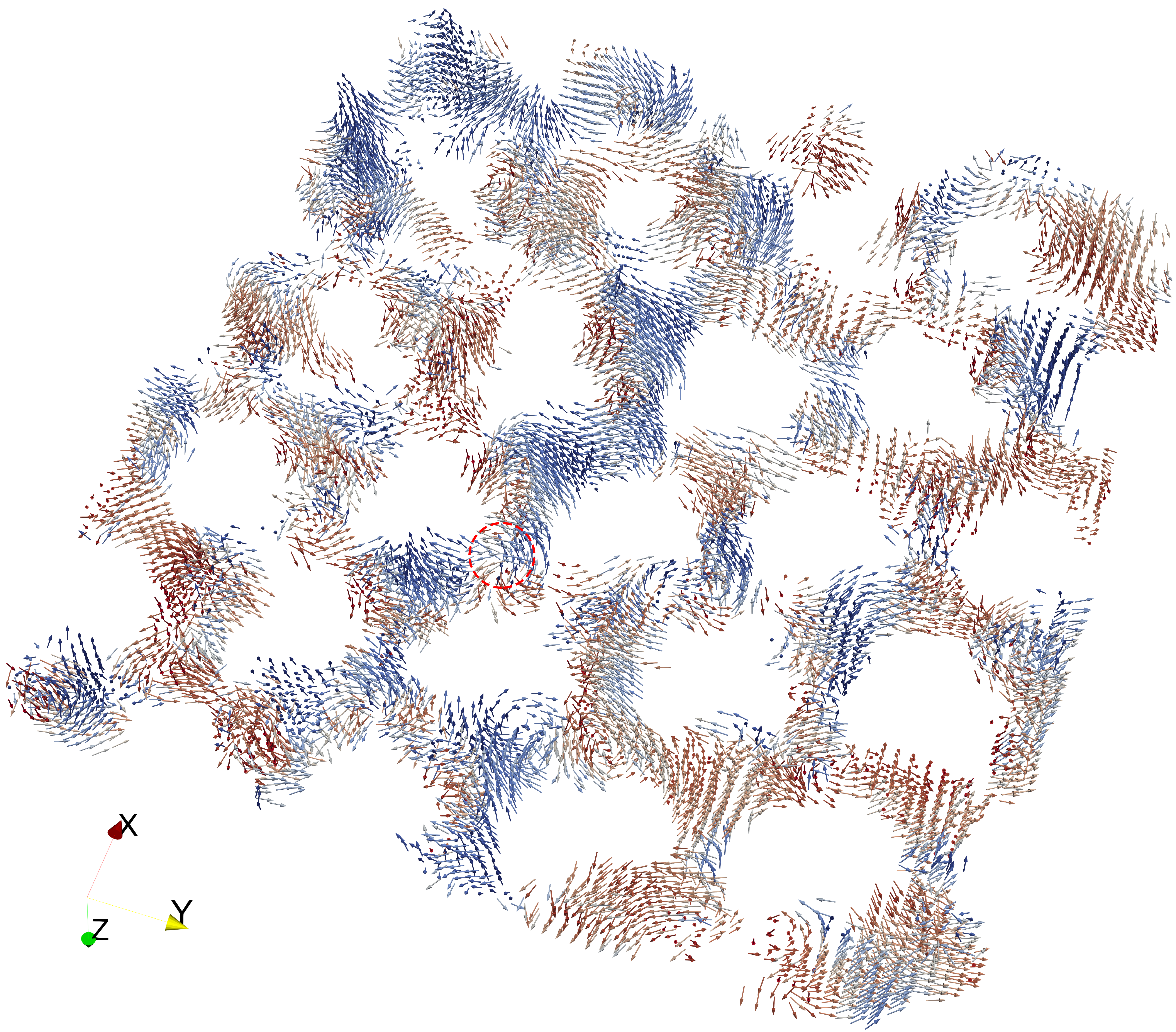}};
            \draw (-2.4, 1.9)node{\textcolor{black}{\textbf{e}}};
        \end{tikzpicture} 
    \end{tabular}
    \caption{3D reconstruction of the magnetization vector filed in a ferromagnetic meta-lattice. (\textbf{a-b}) Representative project with left (\textbf{a}) and right (\textbf{b}) polarization at $0^o$.  (\textbf{c}) The magnetic contrast projection at the $0^o$, which is the difference of the left and right polarization projections. (\textbf{d}) 3D magnetization vector field in the square with dashed lines in (\textbf{a-c}), where the colors represent the different directions of the vectors. (\textbf{e}) A thin layer of (\textbf{d}). (\textbf{f}) A representative topological defect with positive charge. Scale bar 200 nm.}
    \label{metalattice}
\end{figure}

The ferromagnetic meta-lattice consists of 60 nm silica nanoparticles forming a face-centred cubic structure infiltered with nickle.
The vector tomography experiment was conducted at the COSMIC beamline at the Advanced Light Source, Lawrence Berkeley National Lab. 
Circularly polarized x-rays of left- and right-helicity were used to achieve differential magnetic contrast based on x-ray magnetic circular dichrosim\cite{Streubel2015, Chen1990}.
The x-ray energy was set as 856 eV, slightly above the nickel $L_3$ edge
Three in-plane rotation angles (0°, 120° and 240°) were chosen, and tilt ranges is from -62° to +61° for each in-plane rotation angle. 
At each tilt angle, 2D diffraction patterns were reconstructed using the regularized ptychographic iterative engine, producing two projections with left and right polarization at each tilt angle (Fig. \ref{metalattice}a-b).
The scalar tomography reconstruction was performed from three sets of tilt series by summing each pair of the oppositely-polarized projections, from which a 3D support was obtained to separate the magnetic materials from the non-magnetic region. For the vector tomography reconstruction, the difference of the left- and right-circularly polarized projections was calculated (Fig. \ref{metalattice}c), producing magnetic contrast projections of three independent tilt series. 
Using RESIRE-V with the support, we reconstructed the 3D magnetization vector field. Fig. \ref{metalattice}d shows the 3D vector field of the magnified square region with dotted lines in Fig. \ref{metalattice}a-c, where the colors represent the different directions of the vectors. 
A thinner slice of the magnified region and a topological defect with a positive charge are shown in Fig. \ref{metalattice}e-f, respectively. A more detailed analysis of the 3D magnetization vector field and the topological defects in the ferromagnetic meta-lattice can be found elsewhere\cite{rana_2023}. 

\section*{Conclusion}
We present the mathematical formulation and implementation of RESIRE-V, an iterative algorithm for the 3D reconstruction of the vector field. 
RESIRE-V requires the acquisition of multiple tilt series of projections and the algorithm iterates between these projections and a 3D structure by using a forward and a backward step. 
The forward and backward step consist of the Radon transform and a linear transformation, respectively. Our analysis indicates that incorporating a 3D support to separate the magnetic region from a non-magnetic region can help RESIRE-V achieve accurate and robust reconstruction of the 3D vector field. 
To validate RESIRE-V, we perform a numerical simulation of the 3D magnetization vector field in a meta-lattice. Using only two tilt series and a support, we reconstruct the 3D vector field with high accuracy. We also observe that the reconstructed z component has higher quality than the x and y components, which is consistent with our mathematical analysis. 
Finally, we apply RESIRE-V to an experimental data set of a ferromagnetic meta-lattice, consisting of three tilt series with different in-plane rotation angles. Each tilt series has two sets of projections with left and right polarization. 
By using a support constraint, we reconstruct the 3D magnetization vector field inside the ferromagnetic meta-lattice, showing topological defects with positive and negative charges. 
We expect that RESIRE-V can be a general vector tomography method for the 3D reconstruction of a wide range of vector fields


\section*{Supplementary}
\subsection*{Step-size analysis}
After the gradient is established, we need to perform a step-size analysis, showing and proving a step-size that works. The iterations will then be guaranteed to converge to a solution as desired. To proceed with our goal, we approximate the Lipchitz constant L of the gradient $\nabla\varepsilon$, i.e. find L such that the following inequality holds:
\begin{equation}
    \big\| \nabla \varepsilon(\mB_1) - \nabla \varepsilon(\mB_2) \big\| \le L \big\| \mB_1 - \mB_2 \big\| \quad \forall  \; \mB_1, \, \mB_2
\end{equation}
The step-size then can be chosen to be $1/L$ to guarantee a convergence. First, let decompose the error metric function $\varepsilon(\mB)$ into a sum of $N$ sub-error metric functions, which correspond to $N$ projections, i.e. $\varepsilon(\mB) = \sum_{\thetaB} \varepsilon_{\thetaB} ( \mB )$ where
\begin{equation}
    \varepsilon_{\thetaB} ( \mB ) =  \frac{1}{2} { \Big\| \alpha_{\thetaB} \, \Pi_{\thetaB} (M_x) + \beta_{\thetaB} \, \Pi_{\thetaB} (M_y) + \gamma_{\thetaB} \, \Pi_{\thetaB} (M_z)  - b_{\thetaB}^-  \Big\|^2 }
\end{equation}
It suffices to show the Lipchitz constant $L_{\thetaB}$ of the sub-error metric function $\varepsilon_{\thetaB}(\mB)$. The Lipchitz constant $L$ of the sum function $\varepsilon (\mB)$ can be derived via some algebra techniques and triangle inequalities after $L_{\thetaB}$ is obtained. 

Let $N_x\times N_y \times N_z$ be the size of each reconstructed magnetic component $M_x$, $M_y$ and $M_z$. We further assume that $M_x$, $M_y$, $M_z$ and  $\mB$ are vectorized into 1D vectors and that $\mB$ stacks all $M_x$, $M_y$ and $M_z$ together in this respective order. The purpose of this assumption is for matrix analysis only. Hence we can decompose $\big\| \nabla \varepsilon_{\thetaB} (\mB) \big\|^2$ into a sum as follow:
\begin{align}
\big\| \nabla \varepsilon_{\thetaB} (\mB) \big\|^2 = \big\| \frac{\partial \varepsilon_{\thetaB}}{\partial M_y} (\mB) \big\|^2 + \big\| \frac{\partial \varepsilon_{\thetaB}}{\partial M_z} (\mB) \big\|^2  +\big\| \frac{\partial \varepsilon_{\thetaB}}{\partial M_z} (\mB) \big\|^2     
\end{align}
To show an upper bound of $\big\| \frac{\partial \varepsilon_{\thetaB}}{\partial M_x} (\mB) \big\|^2$, we use a sequences of triangle inequalities (or Cauchy–Schwarz):
\begin{align}
    \big\| \frac{\partial \varepsilon_{\thetaB}}{\partial M_x} ( \mB ) \big\|^2 \le \;&  \big\| \alpha_{\thetaB} \big( \alpha_{\thetaB} \Pi_{\thetaB}^T \Pi_{\thetaB} (M_x) + \beta_{\thetaB} \Pi_{\thetaB}^T \Pi_{\thetaB} (M_y) + \gamma_{\thetaB} \Pi_{\thetaB}^T \Pi_{\thetaB} (M_z) \big) \big\|^2 \nonumber\\
    \le \;& 3\, \alpha_{\thetaB}^2 \, \big\| \Pi_{\thetaB}^T \Pi_{\thetaB} \big\|^2 \, \Big( \alpha_{\thetaB}^2  \,  \big\| M_x \big\|^2  +  \beta_{\thetaB}^2 \,  \,  \big\| M_y \big\|^2  +  \gamma_{\thetaB}^2 \,   \big\| M_z \big\|^2 \Big) \nonumber\\
    \le \;& 3 \, \alpha_{\thetaB}^2 \, N_z^2 \max\{ \alpha_{\thetaB}^2, \beta_{\thetaB}^2, \gamma_{\thetaB}^2 \} \, \big\| \mB \big\|^2
\end{align}
Here, we use the result $ \big\| \Pi_{\thetaB}^T \Pi_{\thetaB} \big\| \le N_z $ from elsewhere\cite{Pham_resire}. The Lipchitz constant $L_{\thetaB}$ will be obtained by just summing all three inequality, and taking the square root:
\begin{align}
    \big\|  \nabla \varepsilon_{\thetaB} (\mB_1) - \nabla \varepsilon_{\thetaB} (\mB_2) \big\|   \le  \sqrt{3} \, N_z \, \sqrt{ \alpha_{\thetaB}^2 + \beta_{\thetaB}^2 + \gamma_{\thetaB}^2  } \,  \max\{ |\alpha_{\thetaB}|, |\beta_{\thetaB}|, |\gamma_{\thetaB}| \} \, \big\|  \mB_1 - \mB_2 \big\|
\end{align}    
Using the fact that $\alpha_{\thetaB}^2 + \beta_{\thetaB}^2 + \gamma_{\thetaB}^2 = 1$, we can further simplify $L_{\thetaB} = \sqrt{3} \, N_z  \,  \max\{ |\alpha_{\thetaB}|, |\beta_{\thetaB}|, |\gamma_{\thetaB}| \}$. Note that $\gamma_{\thetaB} = 0 $ for $\theta=0$. We finally obtain an approximation of the Lipchitz constant $L = \sqrt{3} \, n \, N_z$ and the proof is done. Once can choose the step size $t=1/L$ to maximize the convergence speed. In practice, $t$ could be slightly larger than $1/L$ and the algorithm still converges.    
\bibliography{reference}

\section*{Acknowledgements}
This work was supported by 
STROBE: A National Science Foundation Science \& Technology Center under grant number DMR 1548924.
The experimental data has been published elsewhere\cite{rana_2023}, which was conducted at the COSMIC beam line at the Advanced Light Source, Lawrence Berkeley National Lab, which is supported by the Office of Science, Office of Basic Energy Sciences of the US DOE under contract no. DE-AC02-05CH11231.

\section*{Author contributions statement}
J.M. directed the research; M.P. developed the main algorithm with help from X.L. and A.R.; M.P and A.R. performed the numerical simulation and the reconstruction of the experimental data; M.P., J.M. and X.L. wrote the manuscript with contributions from all authors. All authors reviewed the manuscript.

\section*{Data availability statement}
The MATLAB source codes of RESIRE-V and the simulated and experiment data of the meta-lattice are available at the github repository \url{https://github.com/minhpham0309/RESIRE-V}.


\end{document}